\documentclass[sigconf]{acmart}

\usepackage{graphicx}
\usepackage{colortbl} 
\usepackage{xcolor}
\usepackage{array}
\usepackage{enumerate}
\usepackage{xspace}
\usepackage{bbm}
\usepackage{soul}
\usepackage{amstext}
\usepackage{amsmath}
\usepackage{amsthm}
\usepackage{algorithm}
\usepackage{algorithmic}
\usepackage{subfigure}
\usepackage{multirow}
\usepackage{makecell}
\usepackage{comment}

\AtBeginDocument{
  }

\copyrightyear{2025} 
\acmYear{2025} 
\setcopyright{acmlicensed}
\acmConference[WSDM '25]{Proceedings of the
 Eighteenth ACM International Conference on Web Search and Data
 Mining}{March 10--14, 2025}{Hannover, Germany}
 \acmBooktitle{Proceedings of the Eighteenth ACM International Conference on
 Web Search and Data Mining (WSDM '25), March 10--14, 2025, Hannover,
 Germany}
 \acmDOI{10.1145/3701551.3703542}
 \acmISBN{979-8-4007-1329-3/25/03}

\begin{document}

\newcommand{\ours}[0]{Oracle4Rec\xspace}
\newcommand{\name}[0]{oracle-guided dynamic user preference
modeling method for sequential recommendation\xspace}

\title{Oracle-guided Dynamic User Preference Modeling for Sequential Recommendation}

\author{Jiafeng Xia}
\affiliation{
  \institution{Fudan University}
  \city{Shanghai}
  \country{China}
}
\additionalaffiliation{
    \institution{School of Computer Science and Shanghai Key Laboratory of Data Science, Fudan University}
  \city{Shanghai}
  \country{China}
}
\email{jfxia19@fudan.edu.cn}

\author{Dongsheng Li}
\affiliation{
  \institution{Microsoft Research Asia}
  \city{Shanghai}
  \country{China}
 }
 \email{dongsli@microsoft.com}

\author{Hansu Gu}
\affiliation{
  \city{Seattle}
  \country{United States}
}
\email{hansug@acm.org}

\author{Tun Lu}
\authornotemark[1]
\affiliation{
  \institution{Fudan University}
  \city{Shanghai}
  \country{China}
}
\additionalaffiliation{
    \institution{Fudan Institute on Aging, MOE Laboratory for National Development and Intelligent Governance, and Shanghai Institute of Intelligent Electronics \& Systems, Fudan University}
  \city{Shanghai}
  \country{China}
}
\authornote{Corresponding author.}
\email{lutun@fudan.edu.cn}

\author{Peng Zhang}
\authornotemark[1]
\authornotemark[3]
\affiliation{
  \institution{Fudan University}
  \city{Shanghai}
  \country{China}}
\email{zhangpeng_@fudan.edu.cn	}

\author{Li Shang}
\authornotemark[1]
\affiliation{
  \institution{Fudan University}
  \city{Shanghai}
  \country{China}}
\email{lishang@fudan.edu.cn}

\author{Ning Gu}
\authornotemark[1]
\affiliation{
  \institution{Fudan University}
  \city{Shanghai}
  \country{China}}
\email{ninggu@fudan.edu.cn}

\renewcommand{\shortauthors}{Jiafeng Xia et al.}

\begin{abstract}
  Sequential recommendation methods can capture dynamic user preferences from user historical interactions to achieve better performance. However, most existing methods only use past information extracted from user historical interactions to train the models, leading to the deviations of user preference modeling. Besides past information, future information is also available during training, which contains  the ``oracle'' user preferences in the future and will be beneficial to model dynamic user preferences. Therefore, we propose an \name (\ours), which leverages future information to guide model training on past information, aiming to learn ``forward-looking'' models. Specifically, \ours first extracts past and future information through two separate  encoders, then learns a forward-looking model through an oracle-guiding module which minimizes the discrepancy between past and future information. We also tailor a two-phase model training strategy to make the guiding more effective. Extensive experiments demonstrate that \ours is superior to state-of-the-art sequential methods. Further experiments show that \ours can be leveraged as a generic module in other sequential recommendation methods to improve their performance with a considerable margin.
\end{abstract}



\maketitle

\section{Introduction}
Recommender systems, which can recommend potentially interested items to the users according to their historical interactions, have been widely used in various fields, e.g., advertising~\cite{advertising, li2024recommender}, movie recommendation~\cite{moviecf1, moviecf2} and E-commerce~\cite{ec1, ec2, ec3}. Generally, user interactions are continuous and can be seen as a sequence of user's interacted items sorted in chronological order, which makes user preferences constantly change over time in nature. Sequential recommendation methods~\cite{sasrec, bert4rec, fire, fmlprec, igcn, neufilter} can model dynamic user preferences based on his/her historical interaction sequence and achieve better performance compared with static recommendation algorithms~\cite{pmf, autorec, gcmc, widendeep, higsp}.

However, existing sequential recommendation methods are faced with one key challenge. When predicting the next interaction, they only use past information extracted from user interactions that come before the current interaction, which could cause the deviations of dynamic user preference modeling since only using past information is insufficient to capture the dynamics of user preferences~\cite{grec}, and thus hurt the performance of recommendation models. Besides past information, future information, which can be extracted from user interactions that come after the current interaction, is also available in the training. It can be viewed as a type of posterior information containing the ``oracle'' user preferences in the future and is also beneficial to model dynamic user preferences. Using future information to guide model training on past information can learn forward-looking models, which can better model dynamic user preference and thus improve model performance.

To this end, we propose an \name (\ours) that leverages future information to guide model training on past information, aiming to learn forward-looking models. Specifically, \ours first extracts past information and future information through two separate information encoders respectively. The two encoders have the same architecture, which both have a noise filtering module to filter the noise in the user interaction sequence, followed by a causal self-attention module to capture the evolution of user preferences, making the extracted information more accurate. Then \ours learns the forward-looking model through a carefully designed oracle-guiding module in a manner of minimizing the discrepancy between past information and future information. To make the guiding from future information to past information more effective, we also tailor a two-phase model training strategy, named 2PTraining, for \ours. Extensive experiments on six real-world datasets demonstrate that \ours consistently outperforms state-of-the-art sequential recommendation algorithms. We further implement \ours as a generic module due to it is orthogonal to existing sequential methods, and we conduct a generality experiment to show that \ours has high generality and it can be flexibly applied to other sequential methods and improve their performance with a considerable margin. 

The contributions of this work are summarized as follows:
\begin{enumerate}[1.]
    \item[$\bullet$] We propose an \name that can learn forward-looking models through a carefully designed oracle-guiding module and a tailored training strategy, so as to better model dynamic user preferences.
    \item[$\bullet$] We conduct extensive experiments on six public datasets, and the results show that \ours achieves better performance compared with other state-of-the-art sequential methods. 
    \item[$\bullet$] We implement \ours~\footnote{We release the code for further research:~\url{https://github.com/Yaveng/Oracle4Rec}.} as a generic module due to it is orthogonal to existing sequential methods, and experiments show that it can improve the performance of those methods with a considerable margin.
\end{enumerate}

\begin{figure*}[!t]
\centering
\includegraphics[width=0.9\linewidth]{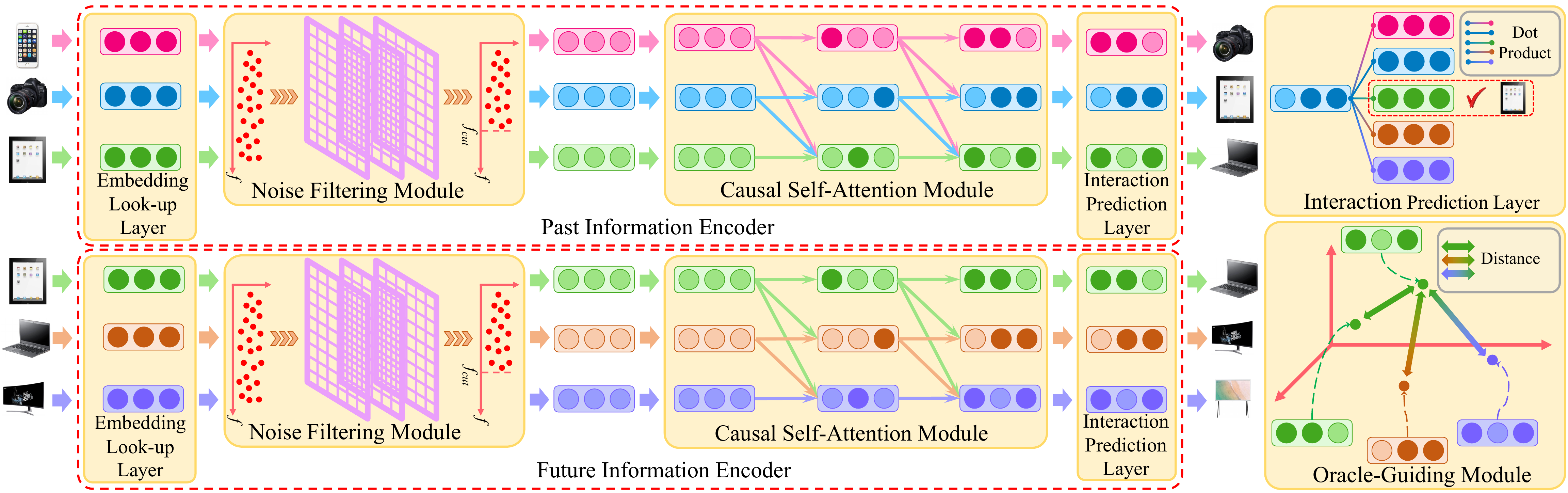}
\caption{The architecture of \ours, which consists of three parts: Past Information Encoder, Future Information Encoder and Oracle-Guiding Module. The first two encoders are both composed of Embedding Look-up Layer, Noise Filtering Module, Causal Self-Attention Module and Interaction Prediction Layer. For the ease of presentation, the Oracle-Guiding Module is set to minimize the distance between past information and future information in a 3D coordinate system.
}
\label{fig:model_stru}
\end{figure*}

\section{The Proposed Method}
\subsection{Notations}
Let user set and item set be $\mathcal{U}$ and $\mathcal{V}$ respectively, $|\mathcal{U}|=m$ and $|\mathcal{V}|=n$. The user $u$'s interaction sequence in chronological order can be denoted as $\mathcal{S}^u=\{v_i| v_i\in \mathcal{V}, i=1,2,\cdots,|\mathcal{S}^u|\}$. Our goal is to predict next item $v_{|\mathcal{S}^u|+1}$ that user $u$ will interact with based on his/her interaction sequence $\mathcal{S}^u$. For the convenience of the following description, we give the definitions of two kinds of user interaction sequences. For a target interaction $v_t\in\mathcal{V}$ that needs to be predicted for user $u$, his/her historical interaction sequence can be defined as $\mathcal{H}^u_t=\{v_1, v_2, \cdots, v_{t-1}\}$, and his/her global interaction sequence is defined as $\mathcal{G}^u_t=\{v_1, v_2, \cdots, v_{t-1}, v_{t}, v_{t+1}, \cdots,  v_{t+P}\}$, which additionally contains $P+1$ items compared with user historical interaction sequence. We usually adopt fixed-length historical interaction sequence and global interaction sequence for high efficiency during training and inference, thus we need to truncate or pad the sequences $\mathcal{H}^u_t$ and $\mathcal{G}^u_t$ to ensure their lengths are equal to a positive integer $L(L>P)$. If the length of sequence is longer than $L$, we truncate its length to $L$ by selecting the recent $L$ interactions, while if the length of sequence is shorter than $L$, we repeatedly add a padding item to the left side of sequence until its length is $L$. Therefore, we redefine the historical interaction sequence as $\mathcal{H}^u_t=\{v_{t-L}, v_{t-L+1}, \cdots, v_{t-1}\}$ and the global interaction sequence as $\mathcal{G}^u_t=\{v_{t+P-L+1}, v_{t+P-L+2}, \cdots, v_{t+P}\}$.

\subsection{Overview of \ours}
\ours is a sequential recommendation method that leverages past information and future information to jointly model dynamic user preferences and make recommendation. Figure~\ref{fig:model_stru} shows the architecture of the proposed \ours, which consists of three key components: a {\bf Past Information Encoder} used to extract past information from user historical interaction sequences, a {\bf Future Information Encoder} used to extract future information from user global interaction sequences, and an {\bf Oracle-Guiding Module} used to guide model training on past information with future information. We also tailor an effective model training strategy 2PTraining to facilitate the guiding from future information to past information, thereby making past information capture the dynamics of user preferences more accurately and sufficiently. Next, we introduce these modules and the tailored training strategy in details.

\subsection{Past Information Encoder}
\subsubsection{Embedding Look-up Layer}
Embedding Look-up Layer is used to generate item embedding matrix $\mathbf{E}^u\in\mathbb{R}^{L\times d}$ 
according to the user historical interaction sequence $\mathcal{H}^u_t$ by looking up the item embedding table $\mathbf{T}\in\mathbb{R}^{|\mathcal{V}|\times d}$. However, unlike GRU~\cite{gru} and LSTM~\cite{lstm}, which can capture the order relation between items in the sequence, the attention mechanism in the following Causal Self-Attention Module will ignore it,  affecting the accuracy of the dynamic user preference modeling. Therefore, we introduce the positional embedding $\mathbf{E}^P\in\mathbb{R}^{L\times d}$ to help the model to identify the order relation between different items. There are many choices of positional embedding, such as sinusoidal positional embedding~\cite{selfattentionmechanism} and learnable positional embedding~\cite{bert}, we follow the suggestion in~\cite{sasrec} to use learnable positional embedding for better performance. Thus, the final item embedding is: 
\begin{equation}
    \hat{\mathbf{E}}^u = \mathbf{E}^u + \mathbf{E}^P.
\label{eq:past_encoder_start}
\end{equation}
Note that we will take user $u$'s historical sequence $\mathcal{H}^u_t$ as an example to illustrate how past information encoder works. Thus, we will omit the superscript of all embeddings that are associated with user $u$ for simplicity hereinafter.

\subsubsection{Noise Filtering Module}
User interaction sequence will inevitably introduce some noisy interactions due to the first impression~\cite{biases}, such as caption bias~\cite{captionbias} and position bias~\cite{locationbias}, which will cause deviations in modeling user preferences. Specifically, the self-attention mechanism employed in the Causal Self-Attention Module (Section~\ref{sec:cau_self_attn_layer}) is highly sensitive to noisy interactions, and the distribution of attention within the model is easily influenced by these noisy interactions, resulting in certain biases in the extracted information. To mitigate the impact of noisy interactions on modeling user dynamic preference, we introduce a Noise Filtering Module before the Causal Self-Attention Module. This adjustment reduces the influence of noisy interactions on modeling user dynamic preference, thereby improving the accuracy of user future interaction predictions.

In the signal processing field, researchers usually analyze and reduce the noise of time series signal in the frequency domain~\cite{fft}. Specifically, the time series signal is first transformed from time domain to frequency domain through Fourier transform, so as to obtain a series of frequency components. Generally, noise corresponds to high-frequency components~\cite{noisehighfreq}. By eliminating a certain proportion of high-frequency components, we can achieve noise reduction of the signal. Then we transform the signal to time domain through inverse Fourier transform, and we obtain a noise-free time series signal. FMLP-Rec~\cite{fmlprec} designs a learnable filter to attenuate the noise in the item embedding by imitating the above process. However, the learnable filter may introduce additional model parameters and increase the difficulty of model training. 

Since noise corresponds to the high-frequency components and our goal is to remove the noise in the item embedding. Thus, we design the Noise Filtering Module to remove high-frequency components in the item embedding to achieve noise reduction. 
The noise filtering module has multiple layers, each layer contains a parameter-free low-pass filter to specify how many low-frequency components are preserved.
Concretely, in the $g$-th layer, we first transform the item embedding $\mathbf{F}^{g-1}$ from time domain to the frequency domain through the Fast Fourier Transform (FFT):
\begin{equation}
    \mathbf{X}^g~, \mathbf{f}^g\leftarrow \text{FFT}(\mathbf{F}^{g-1}),
\end{equation}
where $\mathbf{F}^{g-1}$ is the item embedding from the $(g-1)$-th layer and $\mathbf{F}^0 = \hat{\mathbf{E}}$, $\mathbf{X}^g\in\mathbb{C}^{c\times d}$ and $\mathbf{f}^g\in\mathbb{R}^c$ are the spectrum matrix and frequency vector respectively, $c$ is the number of frequency components. The cutoff frequency $f_{cut}^g$ can be obtained by calculating the $q$ quantile of $\mathbf{f}^g$. Then the indicator matrix $\mathbf{M}^g\in\mathbb{R}^{c\times d}$ , which decides whether the 
frequency components are preserved or removed, can be calculated as:
\begin{equation}   \mathbf{M}^g=\underbrace{\left[\vmathbb{1}_{\mathbf{f}^g_1<f_{cut}^g},\vmathbb{1}_{\mathbf{f}^g_2<f_{cut}^g},\cdots,\vmathbb{1}_{\mathbf{f}^g_c<f_{cut}^g}\right]^T}_{\text{contains }c\text{ elements}} \otimes\mathbf{1}^T, 
\end{equation}
where $\vmathbb{1}$ is an indicator function, $\mathbf{1}$ is a $d$-dimensional vector whose elements are all 1 and $\otimes$ is the outer product. The filtered spectrum $\tilde{\mathbf{X}}^g$ can be obtained by an element-wise product of $\mathbf{X}^g$ and $\mathbf{M}^g$:
\begin{equation}
    \tilde{\mathbf{X}}^g = \mathbf{X}^g\odot \mathbf{M}^g.
\end{equation}
Then we can get the item embedding after noise reduction through the Inverse Fast Fourier Transform (IFFT). Meanwhile, we incorporate the skip connection~\cite{skipconnection}, layer normalization~\cite{layernormalization} and dropout~\cite{dropout} to stabilize training. Thus , the final item embedding $\mathbf{F}^g$ is
\begin{equation}
    \mathbf{F}^g = \text{LayerNorm}(\text{Dropout}(\text{IFFT}(\tilde{\mathbf{X}}^g))+\mathbf{F}^{g-1}).
\end{equation}
After $G$ layers, the output of noise filtering module is $\mathbf{F}$. 

\subsubsection{Causal Self-Attention Module}\label{sec:cau_self_attn_layer}
Compared with RNNs, attention mechanism has surprising long-range dependency modeling ability~\cite{attnlongrange1, attnlongrange2}. Therefore, we use the self-attention mechanism, which borrows from~\cite{sasrec}, to construct the Causal Self-Attention Module, so as to capture the dynamic user preferences after reducing the noise in the item embedding. The causal self-attention module is also composed of multiple layers. In the $k$-th layer, the item embedding $\mathbf{S}^{k-1}$ is first fed into a self-attention layer to capture the global information for all items in the sequence:
\begin{equation}
\begin{split}
    \mathbf{Z}^k &= \text{Softmax}({(\mathbf{S}^{k-1}\mathbf{W}^k_{Q})(\mathbf{S}^{k-1}\mathbf{W}^k_{K})^T}/{\sqrt{d}}),\\
    \tilde{\mathbf{O}}^k &= \mathbf{Z}^k (\mathbf{S}^{k-1}\mathbf{W}^k_{V}),\\
    \mathbf{O}^k &= \text{LayerNorm}(\text{Dropout}(\tilde{\mathbf{O}}^k\mathbf{W}^k_1+\mathbf{b}^k_1)+\mathbf{S}^{k-1}),
\end{split}
\label{eq:self_attn}
\end{equation}
where $\mathbf{W}^k_{Q}$, $\mathbf{W}^k_{K}$ and $\mathbf{W}^k_{V}$ are queries, keys and values projection matrices, and $\mathbf{W}^k_1$ and $\mathbf{b}^k_1$ are weight and bias of linear transformation at $k$-th layer. Skip connection, layer normalization and dropout are used to stabilize training. $\mathbf{S}^{k-1}$ is the item embedding in the $(k-1)$-th layer and $\mathbf{S}^0=\mathbf{F}$. Note that self-attention mechanism does not guarantee the causality that requires the former items cannot receive information from the latter items. Violating the causality will prevent model from learning useful feature and have bad performance in inference. Therefore, when calculating $\mathbf{Z}^k$ in Eq.~(\ref{eq:self_attn}), we should set $\mathbf{Z}^k_{ij}$ to 0 if $i<j (i,j=1,2,\cdots,n)$.

Then we apply a feed-forward network to the item embedding $\mathbf{O}^k$ to capture the non-linearity in the embedding:
\begin{equation}
    \mathbf{H}^{k} = \sigma(\mathbf{O}^k\mathbf{W}^k_2+\mathbf{b}^k_2)\mathbf{W}^k_3+\mathbf{b}^k_3,
\end{equation}
where $\mathbf{W}^k_2$, $\mathbf{W}^k_3$, $\mathbf{b}^k_2$ and $\mathbf{b}^k_3$ are weights and biases, $\sigma(\cdot)$ is the activation function. We also incorporate the skip connection, layer normalization and dropout to stabilize training:
\begin{equation}
    \mathbf{S}^k = \text{LayerNorm}(\text{Dropout}(\mathbf{H}^{k})+\mathbf{O}^{k}).
\end{equation}
After $K$ layers, we denote the output of this module as $\mathbf{S}$. 

\subsubsection{Interaction Prediction Layer}
After obtaining the predicted item embedding $\mathbf{S}$, we can calculate the probability $P(v_{t}|\mathcal{H}^u_t)$ of the target interaction $v_{t}$ based on $\mathcal{H}^u_t$ as:
\begin{equation}
    P(v_{t}|\mathcal{H}^u_t) = \text{Sigmoid}(\mathbf{Q}_{L}^T\mathbf{T}_{v_{t}}),
\end{equation}
where $\mathbf{Q}_{L}$ is the predicted embedding of the target interaction, which corresponds to the last row of $\mathbf{Q}$, $\mathbf{T}_{v_{t}}$ is the real embedding of the target interaction from embedding table $\mathbf{T}$. Therefore, the loss function of past information encoder is:
\begin{equation}
\small\textstyle
    \mathcal{L}_{p}=-\sum_{u=1}^{|\mathcal{U}|}\sum_{t=1}^{|\mathcal{S}^u|}\left[\log(P(v_{t}|\mathcal{H}^u_t))+\log(1-P(j|\mathcal{H}^u_t))\right],\label{eq:past_encoder_end}
\end{equation}
where $\mathcal{S}^u$ is user $u$'s interaction sequence, and $j\notin\mathcal{S}^u$ is a negative item that user $u$ has never interacted with.

\subsection{Future Information Encoder}
Future information encoder is used to encode the user future preferences from user global interaction sequence. It has the same structure as the past information encoder, which consists of embedding look-up layer, noise filtering module, causal self-attention module and interaction prediction layer. Different from BERT4Rec~\cite{bert4rec}, GRec~\cite{grec} and DualRec~\cite{dualrec} that adopt the reversed user interaction sequence to extract future information, i.e., in a Right-to-Left-style, our future information encoder extracts future information in a more natural way, i.e., in a Left-to-Right-style. Given user $u$'s global interaction sequence $\mathcal{G}^u_t$, the future information encoder outputs the item embedding $\mathbf{R}$. Then we can calculate the probability $P(v_{t+P+1}|\mathcal{G}^u_t)$ of the interaction $v_{t+P+1}$: 
\begin{equation}
    P(v_{t+P+1}|\mathcal{G}^u_t) = \text{Sigmoid}(\mathbf{R}_{L}^T\mathbf{T}_{v_{t+P+1}}),
\end{equation}
where $\mathbf{R}_{L}$ is the predicted embedding of item $v_{t+P+1}$ and it is the last row of $\mathbf{R}$, $\mathbf{T}_{v_{t+P+1}}$ is the real embedding of item $v_{t+P+1}$ from embedding table $\mathbf{T}$, and $P$ is the number of future interactions after the target interaction $v_t$ in $\mathcal{G}^u_t$. However, the loss function of future information encoder is slightly different from that of past information encoder. Since we want to incorporate future information to make user preference modeling more accurate, we should ensure that future information is sufficiently extracted from user global interaction sequence. Therefore, we use the following loss function to optimize the future information encoder:
\begin{equation}
\begin{aligned}
\small
\label{eq:fae_loss}
    &\mathcal{L}_{f}=-\sum_{u=1}^{|\mathcal{U}|}\sum_{t=1}^{|\mathcal{S}^u|}\left[\log(P(\mathcal{G}^u_{t+1}|\mathcal{G}^u_t))+\log(1-P(\mathcal{J}|\mathcal{G}^u_t))\right]\\
    &=-\sum_{u=1}^{|\mathcal{U}|}\sum_{t=1}^{|\mathcal{S}^u|}\sum_{l=1}^{L}\left[\log(P(v_{m+l}|\mathcal{G}^u_{t,l}))+\log(1-P(j_l|\mathcal{G}^u_{t,l}))\right],
\end{aligned}
\end{equation}
where $\mathcal{G}^u_{t+1}=\{v_{t+P-L+2}, v_{t+P-L+3}, \cdots, v_{t+P+1}\}$ is user $u$'s global interaction sequence at $t+1$. $\mathcal{J}=\{j_1, j_2, \cdots, j_L\}$ is the negative item sequence containing $L$ negative items that user has never interact with. $\mathcal{G}^u_{t,l}=\{v_m, v_{m+1}, \cdots, v_{m+l-1}\}$ is an sub-sequence of $\mathcal{G}^u_{t}$ and $m=t+P-L+1$.

We share the item embedding and positional embedding between past information encoder and future information encoder, which can bring us two benefits: (1) ensure the consistency of item features between encoders, making the model training stable; (2) reduce the number of trainable parameters for easier model training.

\subsection{Oracle-Guiding Module}
Oracle-Guiding Module is designed to use the future information, as a type of posterior knowledge, to guide the past information to capture the dynamics of user preferences in the future, thereby correcting the deviations in user preference modeling caused by merely using past information. Given two user interaction sequences $\mathcal{H}^u_t$ and $\mathcal{G}^u_t$, we can obtain two item embeddings $\mathbf{Q}$ and $\mathbf{R}$ from past and future information encoder respectively. Our goal is to minimize the discrepancy between the predicted embedding of target interaction $v_t$ from past information encoder, i.e., $\mathbf{Q}_L$ (past information) and the predicted embeddings of the item $v_t$ and the following $P+1$ items from future information encoder, i.e., $\mathbf{R}_{L-P-2+i}~(i=1,2,\cdots,P+2)$ (future information), making the former embedding can capture how user preferences change in the future through the latter embeddings. Thus, the loss function of oracle-guiding module is:
\begin{equation}    
\textstyle
\mathcal{L}_g=\sum_{i=1}^{P+2}\alpha_i\cdot f(\mathbf{Q}_L, \mathbf{R}_{L-P-2+i}),\label{eq:loss_guiding}
\end{equation}
where $\alpha_i\in(0,1]$ is a weight used to distinguish the importance of the $P+2$ discrepancies. Generally, the guiding should pay more attention to the future information near the target interaction $v_t$ and pay less attention to the future information far away from $v_t$, therefore we use exponential attenuation to model $\alpha_i$:
\begin{equation}
    \alpha_i = \text{e}^{-\gamma(i-1)},
\end{equation}
where $\gamma$ is the attenuation coefficient. Since $\mathbf{Q}_L$ and $\mathbf{R}_{L-P-1}$ corresponds to the same item $v_t$, the discrepancy $f(\mathbf{Q}_L, \mathbf{R}_{L-P-1})$ has the highest importance, thus the weight $\alpha_1$ is 1. $f(\cdot)$ is a discrepancy measurement function. There are many choices to model $f(\cdot)$, we introduce four commonly used choices, including \emph{Kullback-Leibler divergence (KL divergence), Jensen–Shannon Divergence (JS Divergence), Euclidean distance and Cosine distance}. Note that for KL divergence and JS Divergence, we need to use the Softmax function to transform $\mathbf{Q}$ and $\mathbf{R}$ to the probabilistic distributions first. We choose KL divergence in \ours since it achieves the best performance in empirical studies.

\subsection{2PTraining: A Tailored Training Strategy}

Traditional training, which jointly trains past and  future information encoder, is unsuitable for \ours since joint training will make the past and future information be extracted simultaneously, weakening the guiding effect of future information. 
Thus, we tailor an effective two-phase model training strategy 2PTraining, which first trains the future information encoder with the loss $\mathcal{L}_f$ to obtain future information, then jointly trains past information encoder and oracle-guiding module with the loss $\mathcal{L}_p+\beta\mathcal{L}_g$ to obtain past information and achieve guiding from future information to past information, to facilitate learning of forward-looking models. We leave the algorithm of 2PTraining in the Appendix due to the space limitation. The experimental results demonstrate the superiority of 2PTraining compared with traditional training strategy.

\subsection{Inference}
In the inference phase, the future information is not available, thus we only use past information encoder to predict the items that users will interact with. Given user $u$'s historical interaction sequence $\mathcal{H}^u_t$, we first use past information encoder to obtain the predicted item embedding $\mathbf{Q}\in\mathbb{R}^{L\times d}$. Since we want to predict which item $v\in\mathcal{V}$ user $u$ will interact with after interacting the item $v_{t-1}$, we can first calculate the interaction score $s^u_{t, v}$, and then the item with the highest score can be seen as the predicted item $v_t$ that user will be interested to interact with:
\begin{equation}
\textstyle
    s^u_{t, v} = \mathbf{Q}_L^T\mathbf{T}_v, \quad
    v_t =\underset{v\in\mathcal{V}}{\arg\max}~s^u_{t,v}.
\end{equation}

\subsection{Discussions}

\subsubsection{Rationality of Adopting Future Information.}
Intuitively, the current prediction can be more accurate when taking future uncertainty into account, since future information, as a kind of hindsight observations, can server as a regularizer to narrow down the representation learning space so as to enhance the accuracy of representation learning and correct the impact of some noisy interactions on user/item representation learning. In deep reinforcement learning, researchers adopt hindsight observation to improve the quality of representation learning~\cite{hindsight1, hindsight2}. For instance, OPD~\cite{hindsight2} trained a teacher network with hindsight information, and employed network distillation to facilitate the learning of student network for stock trading. Inspired by this, we design the oracle-guiding module with the hope that future observations can guide and facilitate the current user preference modeling, thereby more accurately predicting user future interactions.

\subsubsection{Differences Between \ours and Existing Methods.}~\label{sec:diff}
Though \ours, DualRec and GRec adopt past and future information to model dynamic user preferences, there are five major differences between them, and we leave the detailed comparison in the Appendix due to the space limitation:
(1) {\bf The future information extraction manner is different.} DualRec and GRec adopts a right-to-left manner while \ours adopts a left-to-right manner.
(2) {\bf The future information utilization process is different.} DualRec and GRec learns and utilizes future information simultaneously, whereas Oracle4Rec first learns future information and then utilizes it to guide model training.
(3) {\bf The solution to noisy interactions is different.} DualRec and GRec do not consider the impact of noisy interactions on modeling user preference, while Oracle4Rec proposes a lightweight noise filtering module to deal with noisy interactions. 
(4) {\bf The training target for future information encoder is different.} DualRec uses the same loss function for both past and future information encoders, while Oracle4Rec redesigns the future information encoder's loss function (Eq.(\ref{eq:fae_loss})) by considering the guiding role of future information.
(5) {\bf The treatment of future information across periods is different.} DualRec treats future information across periods equally  while Oracle4Rec treats them selectively by assigning them respective weights when narrowing the gap between future and past information.

\subsubsection{Comparison between Embedding Alignment  and Mask \& Prediction.} The Mask \& Prediction technique is commonly used in the NLP domain~\cite{bert} to infer the embedding of a masked word based on its contextual information. BERT4Rec~\cite{bert4rec} inherits this training paradigm and applies it to recommendation algorithms, aiming to infer masked interactions from user interaction sequences. However, we anticipate that current interaction may not always be accurately inferred from future interactions. Therefore, we propose to use the embedding alignment technique, which employs two independent encoders to encode past and future information separately. We use future information to guide model learning on past information. In this training paradigm, our goal is not to infer the current interaction from its context but to regularize the learning of past information using future information, thereby enhancing the consistency of the two types of information. This helps prevent learning spurious correlations, ultimately improving the coherence and accuracy of user preferences modeling. 

Moreover, when user interaction sequence exhibits a degree of randomness and noise, it is more challenging for Mask \& Prediction technique to infer current interaction from future ones. However, encoding user features to capture future changes and guiding model learning on past features are relatively straightforward.

\begin{table}[h!]
\centering
\setlength{\tabcolsep}{2.8pt}
\caption{The statistics of the six real-world datasets.}
\resizebox{1.0\linewidth}{!}{
\begin{tabular}{l|cccccc}
\toprule
&ML100K&ML1M&Beauty&Sports&Toys&Yelp\\
\midrule
\# Users&943&6,040&22,363&25,598&19,412&30,431\\
\# Items&1,349&3,416&12,101&18,357&11,924&20,033\\
\# Interactions&99,287&999,611&198,052&296,337&167,597&316,354\\
Density&7.805\%&4.485\%&0.073\%&0.063\%&0.072\%&0.052\%\\
\bottomrule
\end{tabular}
}
\label{tab:datasets}
\end{table}

\begin{table*}[th!]
  \caption{Performance comparison on six datasets. The best result is denoted in bold, the second best
result is denoted with an underline. The ``$\star$'' denotes the statistical significance ($p$ < 0.05) of the results of \ours compared to the strongest baseline. }
  \label{tab:per_com}
  \resizebox{\linewidth}{!}{
  \begin{tabular}{l|l|ccccccccccccc|c}
    \toprule
Datasets & Metrics & PopRec& GRU4Rec& Caser& RepeatNet& HGN& CLEA& SASRec& BERT4Rec& GRec& SRGNN& GCSAN& FMLP-Rec& DualRec& \ours\\
\midrule
\multirow{6}{*}{ML100K}&HR@1&0.0859&0.1737&0.1680&0.1773&0.1513&0.1741&0.1926&0.1786&0.1669&0.1834&0.2040&0.2085&\underline{0.2097}&{\bf 0.2257}$^\star$\\
&HR@5&0.2397&0.4795&0.4676&0.4732&0.4427&0.4783&0.5005&0.4829&0.4468&0.4806&0.4993&\underline{0.5224}&0.4993&{\bf 0.5330}$^\star$\\
&NDCG@5&0.1619&0.3300&0.3220&0.3296&0.2971&0.3294&0.3499&0.3349&0.3117&0.3370&0.3581&\underline{0.3705}&0.3579&{\bf 0.3843}$^\star$\\
&HR@10&0.3669&0.6566&0.6473&0.6431&0.6036&0.6615&0.6653&0.6361&0.6348&0.6585&0.6664&\underline{0.6783}&0.6670&{\bf 0.6908}$^\star$\\
&NDCG@10&0.2024&0.3872&0.3801&0.3846&0.3514&0.3897&0.4034&0.3846&0.3729&0.3946&0.4121&\underline{0.4208}&0.4108&{\bf 0.4353}$^\star$\\
&MRR&0.1771&0.3205&0.3144&0.3211&0.2891&0.3211&0.3381&0.3243&0.3098&0.3291&0.3489&\underline{0.3562}&0.3481&{\bf 0.3706}$^\star$\\
\midrule
\multirow{6}{*}{ML1M}&HR@1&0.0904&0.3132&0.3119&0.3530&0.2404&0.3203&0.3478&0.3375&0.3448&0.3404&\underline{0.3648}&0.3541&0.3482&{\bf 0.3709}$^\star$\\
&HR@5&0.3002&0.6576&0.6375&0.6547&0.5644&0.6557&0.6803&0.6632&0.6327&0.6474&0.6694&\underline{0.6830}&0.6557&{\bf 0.7106}$^\star$\\
&NDCG@5&0.1964&0.4962&0.4848&0.5138&0.4104&0.5128&0.5255&0.5114&0.4978&0.5034&0.5272&\underline{0.5294}&0.5116&{\bf 0.5531}$^\star$\\
&HR@10&0.4416&0.7754&0.7585&0.7651&0.7087&0.7658&0.7889&0.7716&0.7460&0.7567&0.7773&\underline{0.7916}&0.7727&{\bf 0.8128}$^\star$\\
&NDCG@10&0.2420&0.5345&0.5236&0.5498&0.4572&0.5486&0.5608&0.5467&0.5337&0.5389&0.5623&\underline{0.5648}&0.5496&{\bf 0.5863}$^\star$\\
&MRR&0.2038&0.4689&0.4607&0.4923&0.3922&0.4911&0.4982&0.4857&0.4784&0.4811&\underline{0.5042}&0.5024&0.4901&{\bf 0.5228}$^\star$\\
\midrule
\multirow{6}{*}{Beauty}&HR@1&0.0678&0.1599&0.1304&0.1613&0.1609&0.1327&0.1794&0.1684&0.1317&0.1777&0.1985&0.2038&\underline{0.2182}&{\bf 0.2199}$^\star$\\
&HR@5&0.2105&0.3565&0.3054&0.3287&0.3531&0.3310&0.3880&0.3586&0.2868&0.3650&0.3792&0.4089&\underline{0.4103}&{\bf 0.4317}$^\star$\\
&NDCG@5&0.1391&0.2627&0.2211&0.2482&0.2608&0.2344&0.2886&0.2674&0.2127&0.2753&0.2932&0.3117&\underline{0.3192}&{\bf 0.3316}$^\star$\\
&HR@10&0.3386&0.4596&0.4072&0.4239&0.4544&0.4425&0.4885&0.4598&0.3826&0.4662&0.4707&\underline{0.5079}&0.5029&{\bf 0.5295}$^\star$\\
&NDCG@10&0.1803&0.2960&0.2539&0.2789&0.2934&0.2704&0.3211&0.3000&0.2435&0.3078&0.3227&0.3436&\underline{0.3490}&{\bf 0.3632}$^\star$\\
&MRR&0.1558&0.2633&0.2264&0.2524&0.2616&0.2378&0.2865&0.2689&0.2209&0.2768&0.2940&0.3092&\underline{0.3178}&{\bf 0.3278}$^\star$\\
\midrule
\multirow{6}{*}{Sports}&HR@1&0.0763&0.1291&0.1024&0.1325&0.1444&0.1254&0.1503&0.1405&0.1055&0.1473&0.1674&0.1699&\underline{0.1831}&{\bf 0.1861}$^\star$\\
&HR@5&0.2293&0.3397&0.2834&0.3208&0.3484&0.3397&0.3672&0.3466&0.2786&0.3498&0.3702&0.3915&\underline{0.4031}&{\bf 0.4191}$^\star$\\
&NDCG@5&0.1538&0.2372&0.1947&0.2290&0.2492&0.2327&0.2620&0.2463&0.1937&0.2517&0.2721&0.2846&\underline{0.2981}&{\bf 0.3070}$^\star$\\
&HR@10&0.3423&0.4635&0.4061&0.4381&0.4698&0.4623&0.4964&0.4730&0.3948&0.4738&0.4909&0.5137&\underline{0.5267}&{\bf 0.5431}$^\star$\\
&NDCG@10&0.1902&0.2772&0.2342&0.2668&0.2883&0.2750&0.3037&0.2870&0.2315&0.2916&0.3111&0.3240&\underline{0.3380}&{\bf 0.3471}$^\star$\\
&MRR&0.1660&0.2391&0.2028&0.2342&0.2522&0.2504&0.2632&0.2496&0.2027&0.2552&0.2743&0.2833&\underline{0.2980}&{\bf 0.3037}$^\star$\\
\midrule
\multirow{6}{*}{Toys}&HR@1&0.0585&0.1481&0.1114&0.1370&0.1541&0.1245&0.1799&0.1504&0.1069&0.1682&0.1995&0.1926&{\bf 0.2170}&\underline{0.2152}\\
&HR@5&0.1977&0.3456&0.2944&0.3007&0.3430&0.3278&0.3652&0.3446&0.2687&0.3572&0.3787&0.4007&\underline{0.4043}&{\bf 0.4233}$^\star$\\
&NDCG@5&0.1286&0.2505&0.2054&0.2213&0.2516&0.2286&0.2766&0.2504&0.1891&0.2661&0.2926&0.3015&\underline{0.3149}&{\bf 0.3246}$^\star$\\
&HR@10&0.3008&0.4532&0.4052&0.4007&0.4485&0.4421&0.4609&0.4567&0.3694&0.4605&0.4740&\underline{0.5032}&0.4999&{\bf 0.5222}$^\star$\\
&NDCG@10&0.1618&0.2852&0.2411&0.2535&0.2857&0.2643&0.3075&0.2866&0.2219&0.2995&0.3234&0.3346&\underline{0.3458}&{\bf 0.3565}$^\star$\\
&MRR&0.1430&0.2517&0.2107&0.2279&0.2542&0.2313&0.2778&0.2531&0.1983&0.2680&0.2942&0.2994&\underline{0.3153}&{\bf 0.3215}$^\star$\\
\midrule
\multirow{6}{*}{Yelp}&HR@1&0.0801&0.2109&0.1858&0.2344&0.2337&0.2085&0.2276&0.2446&0.1711&0.2400&0.2561&0.2704&\underline{0.2817}&{\bf 0.2970}$^\star$\\
&HR@5&0.2415&0.5772&0.5170&0.5360&0.5663&0.5678&0.5856&0.5848&0.4743&0.5772&0.5932&0.6231&\underline{0.6272}&{\bf 0.6499}$^\star$\\
&NDCG@5&0.1622&0.3998&0.3560&0.3897&0.4057&0.3842&0.4117&0.4203&0.3251&0.4147&0.4301&0.4538&\underline{0.4610}&{\bf 0.4812}$^\star$\\
&HR@10&0.3609&0.7474&0.6866&0.6950&0.7306&0.7488&0.7651&0.7514&0.6567&0.7398&0.7557&0.7777&\underline{0.7839}&{\bf 0.8000}$^\star$\\
&NDCG@10&0.2007&0.4552&0.4110&0.4411&0.4589&0.4465&0.4699&0.4744&0.3837&0.4675&0.4829&0.5040&\underline{0.5119}&{\bf 0.5300}$^\star$\\
&MRR&0.1740&0.3770&0.3409&0.3778&0.3885&0.3669&0.3915&0.4017&0.3176&0.3966&0.4113&0.4299&\underline{0.4393}&{\bf 0.4562}$^\star$\\
    \bottomrule
  \end{tabular}
  }
\end{table*}

\section{Experiments}
\subsection{Experimental Setup}
We adopt six widely used datasets to comprehensively evaluate the performance of \ours: (1) {\bf ML100K and ML1M} (two movie datasets)~\cite{movielensdataset}. (2) {\bf Beauty, Sports and Toys} (three product datasets)~\cite{amazondataset1, amazondataset2}. (3) {\bf Yelp} (a business dataset)~\cite{yelpdataset}. To make a fair comparison, for all datasets, we group the interaction records by users, and sort them ascendingly in chronological order. Besides, we follow FMLP-Rec to filter unpopular items and inactive users with fewer than five interactions. We also adopt the same dataset split, which is commonly used in sequential recommendation, as FMLP-Rec, i.e., the last item of each user interaction sequence for test, the penultimate item for validation, and all remaining items for training.  Table~\ref{tab:datasets} describes the statistics of the six datasets.

We compare \ours with 13 sequential methods. 
(1) {\bf PopRec}; (2) {\bf GRU4Rec}~\cite{gru4rec}; (3) {\bf Caser}~\cite{caser}; (4) {\bf HGN}~\cite{hgn}; (5) {\bf RepeatNet}~\cite{repeatnet}; (6) {\bf CLEA}~\cite{clea}; (7) {\bf SASRec}~\cite{sasrec}; (8) {\bf BERT4Rec}~\cite{bert4rec}; (9) {\bf GRec}~\cite{grec}; (10) {\bf SRGNN}~\cite{srgnn}; (11) {\bf GCSAN}~\cite{gcsan}; (12) {\bf FMLP-Rec}~\cite{fmlprec}; and (13) {\bf DualRec}~\cite{dualrec}. For all methods, BERT4Rec, GRec and DualRec can access the future information, while other methods can merely use past information for user preference modeling.

We compare \ours with other methods in top-K recommendation task with three popular ranking metrics: (1) {\bf Hit Ratio (HR)}; (2) {\bf Normalized Discounted Cumulative Gain (NDCG)}; and (3) {\bf Mean Reciprocal Rank (MRR)}. For the former two metrics, we report their results when K=1, 5, 10, i.e., HR@\{1, 5, 10\} and NDCG@\{1, 5, 10\}, and we omit NDCG@1 since it is equal to HR@1. For a fair comparison, we follow the same negative sampling strategy as FMLP-Rec to pair the ground-truth item with 99 randomly sampled negative items that the user has not interacted with.

We adopt Adam optimizer~\cite{adam} to optimize the model with learning rates $\eta_1=\eta_2=0.001$, we set the maximum sequence length $L$ to 50 and the number of future interactions $P$ in user global interaction sequence to 10. For all datasets, we tune hidden size $d$ from [32, 64, 128, 256], attenuation coefficient $\gamma$ from 0.005 to 1.0, frequency quantile $q$ from 0.3 to 1.0 and regularization coefficient $\beta$ from [0.005, 0.01, 0.05]. Dropout probabilities are fixed to 0.5. We also tune the layer number of noise filtering module from 1 to 3 and the layer number of causal self-attention module from 1 to 5. The hyper-parameters of all baselines are carefully tuned according to their papers. Due to the space limitation, we leave the detailed setup in the Appendix.

\subsection{Performance Comparison}
Table~\ref{tab:per_com} shows performance comparison of all methods. Following FMLP-Rec, we also report the full-ranking results, i.e., ranking the ground-truth item with all candidate items, in the Appendix. From the results, we have the following observations:

\emph{1. Transformer-based methods (i.e., SASRec and Bert4Rec) achieve better accuracy than RNN-based methods (i.e., GRU4Rec and HGN) and CNN-based methods (i.e., Caser)}. The main reason is that the self-attention mechanism in Transformer-based methods has the largest receptive field compared with RNNs and CNNs, so it can capture more information from user interaction sequence, which can make user preference modeling more precise.

\emph{2. GNN-based methods (i.e., SRGNN and GCSAN) have comparable or better performance than Transformer-based methods}. This is because the GNN has a powerful structure feature extraction ability~\cite{gin}, and can capture the transition relationship between items, so as to achieve more accurate recommendation results.
    
\emph{3. FMLP-Rec and DualRec achieve better accuracy than all other baselines}. This is because the former has strong noise reduction capability to filter the noise in the user interactions, and the latter can utilize both past information and future information from user interactions, making them model user preference more accurate. 

\emph{4. \ours consistently outperforms all compared methods on all datasets, demonstrating its superiority}. Compared with methods only using past information, \ours effectively leverages past and future information in training, making the dynamic user preference modeling more accurate. Compared with BERT4Rec, GRec and DualRec, \ours adopts an oracle-guiding module and a 2PTraining to leverage future information in a more effective way, thus achieving better performance. Generally, using future information can reduce the error of predicting user's interested items during training, and thus will make more accurate recommendation. This can be drawn by comparing the losses, which is placed in the Appendix due to the space limitation, when \ours uses past information only or past and future information together.

\begin{table*}[!hptb]
  \caption{Ablation study of \ours on ML100K, ML1M and Beauty datasets. Bold face indicates the highest performance.}
  \label{tab:abs_stu}
  \resizebox{0.95\linewidth}{!}{
  \small
  \begin{tabular}{l|ccc|ccc|ccc}
    \toprule
    &\multicolumn{3}{c|}{ML100K}&\multicolumn{3}{c|}{ML1M}&\multicolumn{3}{c}{Beauty}\\
    \midrule
    &HR@1&NDCG@5&MRR&HR@1&NDCG@5&MRR&HR@1&NDCG@5&MRR\\
\midrule
(1) \ours w/o Noise Filtering Module&0.1909&0.3554&0.3443&0.3658&0.5495&0.5189&0.2062&0.3104&0.3090\\
(2) \ours w/ Learnable Filter&0.2042&0.3637&0.3513&0.3523&0.5262&0.4995&0.2087&0.3121&0.3110\\
(3) \ours w/o Future Information Encoder&0.1782&0.3391&0.3277&0.3477&0.5275&0.4993&0.1878&0.3010&0.2975\\
(4) \ours w/o Attenuation of Discrepancies&0.1983&0.3604&0.3491&0.3347&0.5225&0.4924&0.2123&0.3255&0.3217\\
(5) \ours w/ Traditional Training Strategy&0.2087&0.3670&0.3566&0.3497&0.5302&0.5016&0.2041&0.3164&0.3129\\
\midrule
(6) \ours w/ JS Diverge&0.1968&0.3583&0.3456&0.3555&0.5399&0.5095&0.2160&0.3257&0.3222\\
(7) \ours w/ Euclidean Distance&0.1987&0.3622&0.3491&0.3630&0.5445&0.5149&0.2071&0.3194&0.3158\\
(8) \ours w/ Cosine Distance&0.2070&0.3712&0.3572&0.3608&0.5439&0.5137&0.2078&0.3211&0.3172\\
\midrule
(9) \ours Future Information Encoder&0.1911&0.3407&0.3328&0.3283&0.5063&0.4798&0.1986&0.3017&0.3004\\
(10) \ours w/ R2L-style &0.2078&0.3659&0.3549&0.3389&0.5157&0.4892&0.2071&0.3174&0.3148\\
\midrule
(11) \ours (w/ KL Divergence \& L2R-style)&{\bf 0.2257}&{\bf 0.3843}&{\bf 0.3706}&{\bf 0.3709}&{\bf 0.5531}&{\bf 0.5228}&{\bf 0.2199}&{\bf 0.3316}&{\bf 0.3278}\\
    \bottomrule
  \end{tabular}
  }
\end{table*}

\begin{table}[t!]
\centering
\caption{The KL divergence between user real and predicted preference under different settings on ML100K and ML1M.}
\resizebox{0.91\linewidth}{!}{
\begin{tabular}{l|c|c}
\toprule
&ML100K&ML1M\\
\midrule
\ours w/o Future Information Encoder&0.0085&0.0069\\
\ours w/ Future Information Encoder &\makecell[c]{0.0070\\(+17.6\%)}&\makecell[c]{0.0062\\(+10.1\%)}\\
\bottomrule
\end{tabular}
}
\label{tab:fie_ana}
\end{table}

\subsection{Ablation Study}\label{sec:aba_stu}
We conduct the extensive ablation study to comprehensively analyze the impact of each component on the performance of \ours. Table~\ref{tab:abs_stu} shows the results. Due to the space limitation, the results of other metrics are placed in the Appendix.

1. Comparing setting (1) and (11), we can find that without the noise filtering module, the performance of \ours significantly decreases on all cases. This is because user interactions are inevitably noisy. By reducing the noise in the interaction sequence using low-pass filter, \ours can more accurately model user preferences, thus making more accurate recommendations. Comparing setting (2) and (11), the results show that low-pass filter (setting (11)) achieves better results than learnable filter, which we attribute to the additional parameters of learnable filter increase the difficulty of model training, thus reducing the performance.
    
2. Comparing setting (3) and (11), we can find that \ours obtains better accuracy than \ours w/o Future Information Encoder, which shows the importance of oracle information guiding in model training, i.e., punish deviations from user future preferences. Moreover, merely using future information encoder can also lead to sub-optimal performance by comparing setting (9) and (11). Therefore, we can conclude that guiding from future information to past information can make user preference modeling more reliable.

3. The attenuation in the oracle-guiding module is helpful to improve the model performance by comparing setting (4) and (11). This is because with attenuation, \ours can selectively minimize the discrepancies between past and future information from different time, thus capturing the evolution of user preferences and achieving better performance.  

4. We can find that 2PTraining is beneficial to improve the performance of \ours by comparing setting (5) and (11). Traditional training strategy jointly trains past and future information encoders, which is unable to sufficiently leverage past information and future information. However, 2PTraining first optimizes future information encoder then optimizes the past information encoder, making the model able to fully leverage these two information. 

5. Comparing setting (6)--(8) and (11), we can find that the performance of \ours varies with different discrepancy measurement methods.  When equipped with KL divergence, \ours achieves the best results. Thus, we adopt KL divergence as the discrepancy measurement method in \ours.

6. The L2R-style (left-to-right-style) information extraction manner is superior than the R2L-style (right-to-left-style) information extraction manner by comparing setting (10) and (11), which demonstrates the rationality of our left-to-right-style. Generally, the left-to-right-style is more intuitive and aligns with the chronological order of interactions, making the extracted information more sufficient and the modeled user preference more accurate.

\subsection{Future Information Encoder Analysis}
To explore how future information encoder can facilitate the learning process of \ours, or in other words, what \ours will learn when equipped with future information encoder, we analyze the consistency of user real preference and user predicted preference on ML100K and ML1M. The former preference is calculated from the categories of real items that user interacts with in the inference phase, which is denoted as $p$, and the latter preference is calculated from the categories of top-10 items predicted by \ours w/o future information encoder and \ours w/ future information encoder, denoted as $q^{(1)}$ and $q^{(2)}$. By comparing the KL divergence $\mathbf{KL}(p, q^{(1)})$ and $\mathbf{KL}(p, q^{(2)})$, we can determine whether future information encoder benefits the learning process of \ours. We leave the details about the calculation of preference and KL divergence in the Appendix due to the space limitation.

Table~\ref{tab:fie_ana} shows the experimental results. We can observe that the KL divergence is reduced when \ours is equipped with future information encoder, which demonstrates that future information encoder can make predicted items more similar to items that user will interact with in terms of item categories and is beneficial to the leaning process of \ours. Generally, future information encoder can help infer ``oracle'' user preference in the future, and with the help of oracle-guiding module, \ours can recognize how user preference will change from past to future, which cannot be achieved by merely using past information encoder, thus making user dynamic preference modeling more accurate.

\begin{figure}[t!]
\centering
\includegraphics[width=1.0\linewidth]{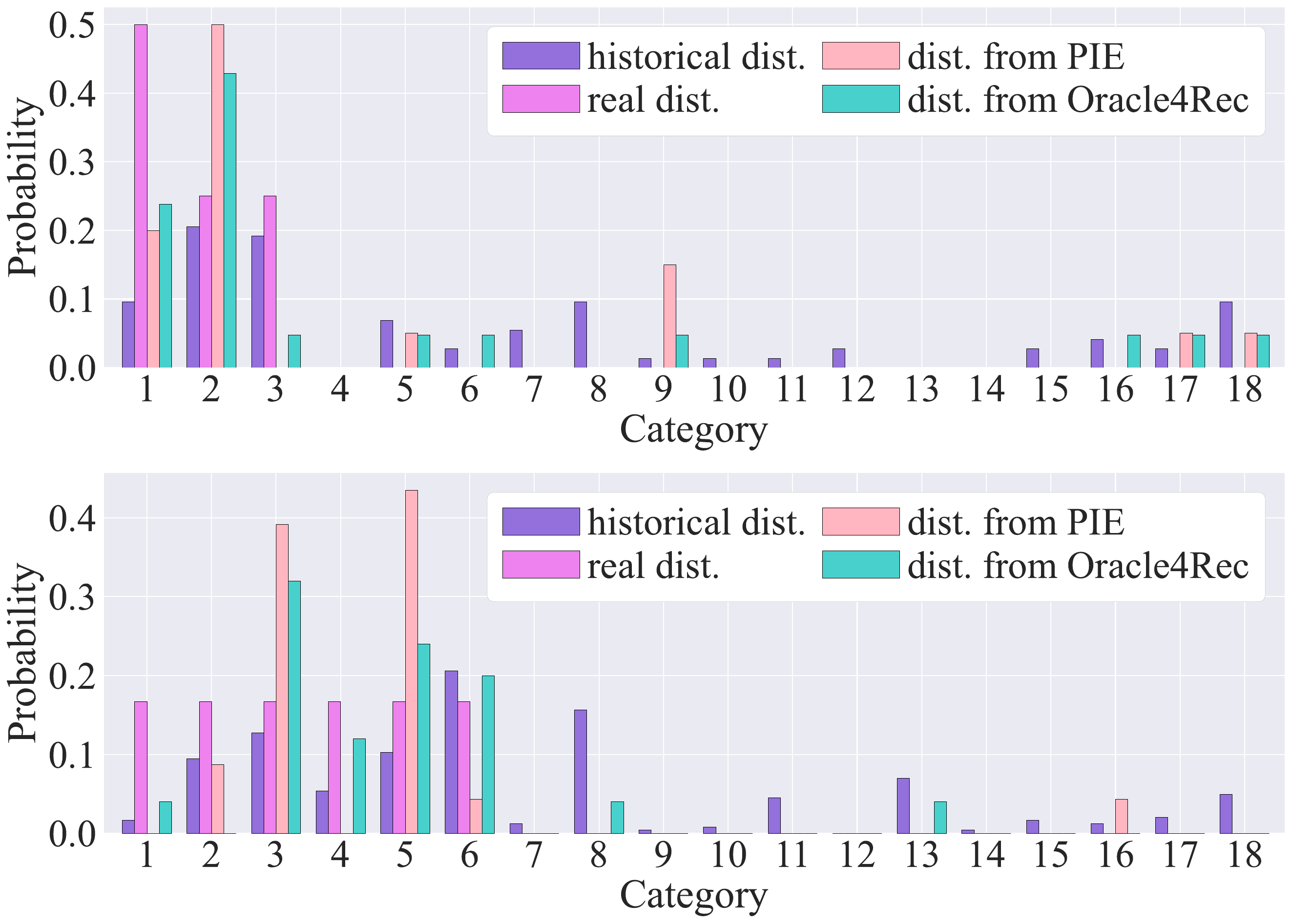}
\caption{Four preference distributions of user 260 (the upper figure) and 869 (the lower figure) on ML100K. ``dist.'' means distribution, and ``PIE'' is past information encoder, i.e., \ours w/o Future information Encoder.}
\label{fig:dist_case}
\end{figure}

Figure~\ref{fig:dist_case} shows four preference distributions of two randomly selected users on ML100K. More cases are placed in the Appendix. Note that historical distribution is calculated from the categories of items that user has interacted with in the training phase. From the results, we can find that merely using past information will make the model easily overfit historical preference. For instance, items related to category 2 have higher probability to be recommended to user 260 than other categories, making the model overfit this category by comparing historical distribution and distribution from PIE. Same phenomenon can be found in the category 3 and 5 of user 869. However, with future information encoder, the probabilities of those categories are reduced by comparing distribution from PIE and distribution from \ours, alleviating the overfiting issue. Moreover, future information encoder makes the predicted preference distribution closer to the real distribution by comparing real distribution, distribution from PIE and distribution from \ours, e.g., category 1 and 2 for user 260 and category 3---6 for user 869, leading to more accurate user interaction prediction.

\subsection{Generality Analysis}
We apply the oracle-guiding module and 2PTraining of \ours into four sequential methods GRU4Rec, SRGNN, GCSAN and FMLP-Rec and evaluate their performance afterwards to further analyze the generality of \ours. Table~\ref{tab:gen_ana} shows the results, the results of other metrics are placed in the Appendix. We can find that the performance of all methods has a significant improvement, which shows \ours has high generality. The results also show that leveraging future information can improve the performance of sequential methods, since future information as a type of posterior knowledge contains the evolution of user preferences in the future, which can address the deviations caused by only using past information to model dynamic user preferences as GRU4Rec, SRGNN, GCSAN and FMLP-Rec do, thus achieving superior performance.

\begin{table}[t!]
  \caption{Generality analysis of four sequential methods. ``XYZ+'' means applying \ours into the corresponding method XYZ. Bold face indicates better performance in that group. The ``$\star$'' denotes the statistical significance ($p$ < 0.05) of the results of XYZ+ compared to XYZ.}
  \label{tab:gen_ana}
  \resizebox{1.0\linewidth}{!}{
  \begin{tabular}{l|ccc|ccc}
    \toprule
    &\multicolumn{3}{c|}{ML100K}&\multicolumn{3}{c}{ML1M}\\
    \midrule
    &HR@1&NDCG@5&MRR&HR@1&NDCG@5&MRR\\
\midrule
GRU4Rec&0.1737&0.3300&0.3205&0.3132&0.4962&0.4689\\
GRU4Rec+&{\bf 0.2059}$^\star$&{\bf 0.3642}$^\star$&{\bf 0.3532}$^\star$&{\bf 0.3446}$^\star$&{\bf 0.5244}$^\star$&{\bf 0.4964}$^\star$\\
\midrule
SRGNN&0.1834&0.3370&0.3291&0.3404&0.5034&0.4811\\
SRGNN+&{\bf 0.1985}$^\star$&{\bf 0.3574}$^\star$&{\bf 0.3466}$^\star$&{\bf 0.3578}$^\star$&{\bf 0.5244}$^\star$&{\bf 0.5006}$^\star$\\
\midrule
GCSAN&0.2040&0.3581&0.3489&0.3648&0.5272&0.5042\\
GCSAN+&{\bf 0.2159}$^\star$&{\bf 0.3741}$^\star$&{\bf 0.3613}$^\star$&{\bf 0.3846}$^\star$&{\bf 0.5462}$^\star$&{\bf 0.5231}$^\star$\\
\midrule
FMLP-Rec&0.2085&0.3705&0.3562&0.3541&0.5294&0.5024\\
FMLP-Rec+&{\bf 0.2178}$^\star$&{\bf 0.3788}$^\star$&{\bf 0.3658}$^\star$&{\bf 0.3657}$^\star$&{\bf 0.5396}$^\star$&{\bf 0.5125}$^\star$\\

    \bottomrule
  \end{tabular}
 }
\end{table}

\subsection{Sensitivity Analysis}
We analyse how \ours performs with respect to three important hyper-parameters number of future interactions, frequency quantile and attenuation coefficient on Beauty. However, due to the space constraints, we leave experimental results in the Appendix. 

\section{Related Work}
Sequential recommendation methods model dynamic user preferences according to his/her historical interactions~\cite{hgn, repeatnet, clea}. Several techniques have been applied in this line of work, from the earliest Markov Chain~\cite{fpmc, fossil},  to the later Recurrent Neural Networks~\cite{rrn, gru4rec}, Convolutional Neural Networks~\cite{3dcnn, caser}, Self-Attention Mechanism~\cite{sasrec, bert4rec} and Graph Neural Networks~\cite{srgnn, gcsan}. FPMC~\cite{fpmc} is a Markov Chain-based sequential method that combines Markov chain and matrix factorization to model user preferences. Fossil~\cite{fossil} fuses the similarity-based method with Markov Chain to characterize users in terms of both preferences and the strength of sequential behavior, addressing the sparsity issues and the long-tailed distribution of datasets. Due to the vigorous development of deep learning, the performance of sequential recommendation methods has been significantly improved. 
GRU4Rec~\cite{gru4rec} adopts GRU~\cite{gru} to model the dynamic user preferences. 3D-CNN~\cite{3dcnn} uses a convolutional neural network to capture sequential patterns and model user preference. 
SASRec~\cite{sasrec} uses a self-attention mechanism~\cite{selfattentionmechanism} to capture user long-term preferences. 
SRGNN~\cite{srgnn} and GCSAN~\cite{gcsan} are GNN-based sequential methods that use graph neural networks~\cite{gnn, gcn} and self-attention mechanism to model user preferences. FMLP-Rec~\cite{fmlprec} adopts an all-MLP architecture to model dynamic user preferences. Note that contrastive learning based sequential methods are another line of works, they are different from ours, thus we exclude them.

However,  the above methods only use the past information to make recommendation, making the model performance sub-optimal. Future information is also available during training and is beneficial to model user preferences. BERT4Rec~\cite{bert4rec}, GRec~\cite{grec} and DualRec~\cite{dualrec} are three sequential methods that integrates past and future information to model dynamic user preferences. Though \ours also leverages future information to model dynamic user preferences, there are several major differences between existing methods and \ours, as discussed in Section~\ref{sec:diff}.

\section{Conclusion}
We propose an \name that uses future information to guide model training on past information, so as to better model dynamic user preferences. \ours first extracts past and future information through two separate encoders, then learns forward-looking models through an oracle-guiding module and a tailored model training strategy. 
Extensive experimental results demonstrate the superiority of \ours. The generality experiment further shows that \ours can be flexibly applied to other sequential methods and greatly improve their performance. 

\begin{acks}
This research was supported by National Natural Science Foundation of China (NSFC) under the Grant No. 62172106, 61932007, and 62372113.
\end{acks}

\section*{Ethical Considerations}
To the best of our knowledge, this work does not bring new negative societal impacts, including fairness, privacy, security, safety, misuse of the technology by malicious actors, as well as possible harms that could arise even when the technology is being used as intended and functioning correctly.

\bibliographystyle{ACM-Reference-Format}
\bibliography{main}

\clearpage
\appendix

\section{Additional Details of \ours}
\subsection{The Algorithm of 2PTraining}
 Algorithm~\ref{alg:2ptraining} shows the workflow of the 2PTraining. At first, 2PTraining randomly initializes model parameters (line 1). Then for each epoch, 2PTraining first trains the future information encoder (line 3--4), followed by a forward propagation to obtain the sufficiently encoded future information (line 5), and then trains the past information encoder and oracle-guiding module to achieve the guiding from future information to past information (line 6--8).
 
\begin{algorithm}[t!]
\caption{The workflow of 2PTraining.}
\label{alg:2ptraining}
\begin{flushleft}
\textbf{Input}: User historical interaction sequence $\mathcal{H}^u_t$, user global interaction sequence $\mathcal{G}^u_t$, learning rates $\eta_1$ and $\eta_2$, number of training epochs $I$, regularization coefficient $\beta$.\\
\textbf{Parameters}: Parameters of past information encoder $\mathbf{\Theta}_p$, parameters of future information encoder $\mathbf{\Theta}_f$. 
\\
\end{flushleft}
\begin{algorithmic}[1]
\STATE Randomly initialize $\mathbf{\Theta}_p$ and $\mathbf{\Theta}_f$. 
\FOR{$i=1, \cdots, I$}
\STATE Train the future information encoder with $\mathcal{G}^u_t$  to obtain $\mathcal{L}_f$.\\
\STATE Update parameters $\mathbf{\Theta}_f$ using gradient descent:\\$\mathbf{\Theta}_f\leftarrow \mathbf{\Theta}_f-\eta_1\frac{\partial \mathcal{L}_f}{\partial \mathbf{\Theta}_f}$.\\
\STATE Feed forward $\mathcal{G}^u_t$  into the future information encoder once and obtain future information $\mathbf{R}$
\STATE Train the past information encoder with $\mathcal{H}^u_t$ to obtain past information $\mathbf{Q}$ and loss $\mathcal{L}_p$.\\
\STATE Train the Oracle-Guiding Module with $\mathbf{Q}$ and $\mathbf{R}$ and obtain the loss $\mathcal{L}_g$.\\
\STATE Update parameters $\mathbf{\Theta}_p$ using gradient descent:\\ $\mathbf{\Theta}_p\leftarrow \mathbf{\Theta}_p-\eta_2\frac{\partial(\mathcal{L}_p+\beta\cdot\mathcal{L}_g)}{\partial \mathbf{\Theta}_p}$.\\ 

\ENDFOR
\end{algorithmic}
\end{algorithm}

\subsection{The detailed differences between \ours and existing methods}
Though both Oracle4Rec, DualRec and GRec adopt past information and future information to model dynamic user preferences, there are five major differences between them:

1. DualRec and GRec extract future information from user interaction sequences in a right-to-left manner, while Oracle4Rec extracts future information in a left-to-right manner. The latter is more intuitive, aligning with the chronological order of interactions, where the usage of items should follow the sequence of interactions.

2. DualRec and GRec employ joint training of past and future information encoders to learn user dynamic preferences, whereas Oracle4Rec first trains a future information encoder to encode future information and then trains a past encoder to guide the model training through future information. We believe that simultaneously training both encoders in the former may lead to mutual interference between them, causing instability and inadequacy in feature learning. In contrast, the two-stage training approach in the latter can avoid mutual interference, resulting in more stable feature learning and more effective utilization of future information.

3. DualRec does not consider the impact of noisy interactions on modeling user dynamic preferences, while Oracle4Rec addresses the issue of widespread noise in user interaction sequences by introducing a lightweight (no trainable parameters) noise filtering module. This module reduces the impact of noisy interactions on modeling user preferences without adding a burden to the model training.

4. DualRec uses the same loss function for both past and future information encoders, extracting past and future information from user interaction sequences. In contrast, Oracle4Rec, considering the guiding role of future information, redesigns the encoder's loss function (Equation 12). This ensures accurate and comprehensive encoding of future information from user interaction sequences, guiding model learning on past information and ensuring the coherence of user preference changes.

5. DualRec treats future information from different time equally when narrowing the gap between future and past information, without considering the varying importance of future information at different time. In Oracle4Rec, the narrowing of the gap is selectively applied. Specifically, future information closer to the current time has a more significant impact on guidance at the current time. Therefore, larger weights should be assigned when reducing the gap. In contrast, future information from the current time has a less significant impact, so smaller weights should be assigned when narrowing the gap. This selective approach in Oracle4Rec results in user preferences with higher coherence and accuracy.

\section{Additional Details of experimental Setup}\label{sec:setup_app}

\subsection{Datasets} We adopt six widely used datasets to comprehensively evaluate the performance of \ours: (1) {\bf ML100K and ML1M} are two popular movie recommendation datasets collected by GroupLens Research from the MovieLens web site. (2) {\bf Beauty, Sports and Toys} are three product recommendation datasets collected from Amazon.com. (3) {\bf Yelp} is a business recommendation dataset. We only use the user interaction records after January 1st, 2019 since it is a very large dataset.

To make a fair comparison, for all datasets, we group the interaction records by users, and sort them ascendingly in chronological order. Besides, we follow~\cite{hidasi2016parallel,fpmc} to filter unpopular items and inactive users with fewer than five interaction records. We also adopt the same dataset split as FMLP-Rec~\cite{fmlprec}, i.e., the last item of each user interaction sequence for test, the penultimate item for validation, and all remaining items for training. 

\subsection{Compared Methods}
We compare \ours with the following 13 sequential recommendation methods. 
\begin{enumerate}[1.]
    \item[$\bullet$] {\bf PopRec} is a simple method that ranks items based on the item popularity.
    
    \item[$\bullet$] {\bf GRU4Rec}~\cite{gru4rec} uses GRU to model dynamic user preferences.
    
    \item[$\bullet$] {\bf Caser}~\cite{caser} adopts horizontal and vertical convolutions to model user short-term preferences.
    
    \item[$\bullet$] {\bf HGN}~\cite{hgn} integrates a hierarchical gating network with the BPR~\cite{bpr} to capture both  long-term and short-term user interests.
    
    \item[$\bullet$] {\bf RepeatNet}~\cite{repeatnet} uses an encoder-decoder structure that can choose items from a user’s history and recommends them at the right time through a repeat recommendation mechanism.
    
    \item[$\bullet$] {\bf CLEA}~\cite{clea}  uses contrastive learning technique to automatically extract items relevant to the target item for recommendation.
    
    \item[$\bullet$] {\bf SASRec}~\cite{sasrec} uses a self-attention mechanism to capture user long-term preferences based on relatively few actions. 
    
    \item[$\bullet$] {\bf BERT4Rec}~\cite{bert4rec} employs a bidirectional self-attention to model user behavior sequences.
    
    \item[$\bullet$] {\bf GRec}~\cite{grec} integrates past information and future information through a gap-filling mechanism.
    
    \item[$\bullet$] {\bf SRGNN}~\cite{srgnn}  models user interaction sequences as a session graph and uses GNN and Attention Network to capture dynamic user preferences.
    
    \item[$\bullet$] {\bf GCSAN}~\cite{gcsan} is a state-of-the-art method that uses GNN and Self-Attention Mechanism to model dynamic user preferences over a session graph generated from user interaction sequences.
    
    \item[$\bullet$] {\bf FMLP-Rec}~\cite{fmlprec} is a state-of-the-art method that adopts an all-MLP structure with learnable filters to filter the noise in user interaction sequences and model accurate user preferences.
    
    \item[$\bullet$] {\bf DualRec}~\cite{dualrec} proposed a dual network to achieve past-future disentanglement and past-future mutual enhancement, so as to alleviate the training-inference gap.
\end{enumerate}

\subsection{Metrics}
We compare \ours with other state-of-the-art sequential recommendation methods in top-K recommendation task with three kinds of popular ranking metrics: (1) {\bf Hit Ratio (HR)}, which evaluates the coincidence of user recommendation list and ground-truth interaction list; (2) {\bf Normalized Discounted Cumulative Gain (NDCG)}, which accumulates the gains from ranking list with the discounted gains at lower ranks; (3) {\bf Mean Reciprocal Rank (MRR)}, which evaluates the performance of ranking according to the harmonic mean of the ranks. Note that for the former two metrics, we report their results when K=1, 5, 10, that is HR@\{1, 5, 10\} and NDCG@\{1, 5, 10\}, and we omit NDCG@1 metric since it is equal to HR@1. For a fair comparison, we follow the same negative sampling strategy as FMLP-Rec~\cite{fmlprec} to pair the ground-truth item with 99 randomly sampled negative items that the user has not interacted with. 

{\color{red}
\begin{table*}[th!]
\centering
  \caption{Performance comparison on six datasets. The best result is denoted in bold, the second best
result is denoted with an underline. The ``$\star$'' denotes the statistical significance ($p$ < 0.05) of the results of \ours compared to the strongest baseline. }
  \label{tab:per_com_full}
  \resizebox{0.88\linewidth}{!}{
  \begin{tabular}{l|l|cccccccc|c}
    \toprule
Datasets & Metrics & GRU4Rec& RepeatNet& HGN& SASRec& BERT4Rec& GCSAN& FMLP-Rec& DualRec& \ours\\
\midrule
\multirow{6}{*}{ML100K}&HR@1&0.0993&0.1046&0.0855&0.0886&0.1101&\underline{0.1241}&0.1184&0.1214&{\bf 0.1317}$^\star$\\
&HR@5&0.0619&0.0677&0.0527&0.0695&0.0710&\underline{0.0794}&0.0753&0.0792&{\bf 0.0867}$^\star$\\
&NDCG@5&0.1637&0.1784&0.1427&0.1767&0.1771&\underline{0.1968}&0.1930&0.1833&{\bf 0.2083}$^\star$\\
&HR@10&0.0826&0.0911&0.0711&0.0917&0.0926&\underline{0.1026}&0.0992&0.0972&{\bf 0.1112}$^\star$\\
&NDCG@10&0.2623&0.2785&0.2195&0.2762&0.2825&\underline{0.2952}&0.2471&0.2661&{\bf 0.3086}$^\star$\\
&MRR&0.1074&0.1162&0.0903&0.1166&0.1190&\underline{0.1274}&0.1134&0.1194&{\bf 0.1364}$^\star$\\
\midrule
\multirow{6}{*}{ML1M}&HR@1&0.1116&0.1311&0.0817&0.1339&0.1331&\underline{0.1565}&0.1361&0.1470&{\bf 0.1613}$^\star$\\
&HR@5&0.0705&0.0857&0.0512&0.0865&0.0851&\underline{0.1026}&0.0884&0.1004&{\bf 0.1062}$^\star$\\
&NDCG@5&0.1864&0.2034&0.1348&0.2106&0.2105&\underline{0.2330}&0.2153&0.2116&{\bf 0.2456}$^\star$\\
&HR@10&0.0945&0.1088&0.0682&0.1111&0.1099&\underline{0.1275}&0.1138&0.1212&{\bf 0.1334}$^\star$\\
&NDCG@10&0.2894&0.2863&0.2114&0.3147&0.3152&\underline{0.3249}&0.2735&0.2956&{\bf 0.3569}$^\star$\\
&MRR&0.1204&0.1314&0.0874&0.1373&0.1363&\underline{0.1512}&0.1292&0.1423&{\bf 0.1614}$^\star$\\
\midrule
\multirow{6}{*}{Beauty}&HR@1&0.0183&0.0365&0.0321&0.0361&0.0324&\underline{0.0451}&0.0403&0.0403&{\bf 0.0526}$^\star$\\
&HR@5&0.0113&0.0255&0.0203&0.0232&0.0209&\underline{0.0311}&0.0260&0.0287&{\bf 0.0347}$^\star$\\
&NDCG@5&0.0329&0.0537&0.0528&0.0561&0.0527&\underline{0.0653}&\underline{0.0653}&0.0556&{\bf 0.0784}$^\star$\\
&HR@10&0.0159&0.0311&0.0269&0.0296&0.0274&\underline{0.0376}&0.0340&0.0336&{\bf 0.0430}$^\star$\\
&NDCG@10&0.0564&0.0777&0.0817&0.0837&0.0794&\underline{0.0927}&0.0838&0.0772&{\bf 0.1132}$^\star$\\
&MRR&0.0202&0.0371&0.0343&0.0366&0.0341&\underline{0.0445}&0.0389&0.0390&{\bf 0.0517}$^\star$\\
\midrule
\multirow{6}{*}{Sports}&HR@1&0.0110&0.0174&0.0176&0.0187&0.0156&0.0227&0.0247&\underline{0.0252}&{\bf 0.0285}$^\star$\\
&HR@5&0.0071&0.0116&0.0112&0.0124&0.0098&0.0154&0.0162&\underline{0.0168}&{\bf 0.0186}$^\star$\\
&NDCG@5&0.0181&0.0277&0.0296&0.0294&0.0257&0.0345&0.0379&\underline{0.0394}&{\bf 0.0440}$^\star$\\
&HR@10&0.0094&0.0149&0.0150&0.0158&0.0130&0.0192&0.0204&\underline{0.0212}&{\bf 0.0236}$^\star$\\
&NDCG@10&0.0300&0.0424&0.0464&0.0449&0.0414&0.0505&0.0479&\underline{0.0600}&{\bf 0.0665}$^\star$\\
&MRR&0.0124&0.0186&0.0192&0.0197&0.0169&0.0232&0.0231&\underline{0.0265}&{\bf 0.0293}$^\star$\\
\midrule
\multirow{6}{*}{Toys}&HR@1&0.0217&0.0333&0.0303&0.0462&0.0305&\underline{0.0573}&0.0526&0.0561&{\bf 0.0625}$^\star$\\
&HR@5&0.0142&0.0245&0.0206&0.0312&0.0205&\underline{0.0419}&0.0355&0.0394&{\bf 0.0420}\\
&NDCG@5&0.0352&0.0466&0.0479&0.0676&0.0476&0.0784&0.0774&\underline{0.0794}&{\bf 0.0891}$^\star$\\
&HR@10&0.0185&0.0288&0.0262&0.0381&0.0260&\underline{0.0486}&0.0435&0.0470&{\bf 0.0506}$^\star$\\
&NDCG@10&0.0558&0.0643&0.0717&0.0931&0.0710&0.1043&0.0947&\underline{0.1083}&{\bf 0.1223}$^\star$\\
&MRR&0.0237&0.0332&0.0322&0.0445&0.0320&\underline{0.0552}&0.0480&0.0542&{\bf 0.0589}$^\star$\\
\midrule
\multirow{6}{*}{Yelp}&HR@1&0.0120&0.0155&0.0151&0.0150&0.0163&\underline{0.0211}&0.0182&\underline{0.0211}&{\bf 0.0233}$^\star$\\
&HR@5&0.0076&0.0097&0.0094&0.0094&0.0099&0.0132&0.0113&\underline{0.0134}&{\bf 0.0147}$^\star$\\
&NDCG@5&0.0211&0.0268&0.0263&0.0256&0.0292&0.0350&0.0314&\underline{0.0357}&{\bf 0.0393}$^\star$\\
&HR@10&0.0105&0.0133&0.0130&0.0128&0.0141&0.0177&0.0155&\underline{0.0180}&{\bf 0.0198}$^\star$\\
&NDCG@10&0.0364&0.0452&0.0446&0.0423&0.0496&0.0575&0.0429&\underline{0.0584}&{\bf 0.0655}$^\star$\\
&MRR&0.0143&0.0179&0.0176&0.0170&0.0191&0.0233&0.0186&\underline{0.0235}&{\bf 0.0263}$^\star$\\

    \bottomrule
  \end{tabular}
  }
\end{table*}
}

\subsection{Implementation Details}
We implement our method using PyTorch. We adopt Adam optimizer~\cite{adam} to optimize the model with learning rates $\eta_1=\eta_2=0.001$, we set the maximum sequence length $L$ to 50 and the number of future interactions $P$ in user global interaction sequence to 10. For all datasets, we tune hidden size $d$ from [32, 64, 128, 256], attenuation coefficient $\gamma$ from 0.005 to 1.0, frequency quantile $q$ from 0.3 to 1.0 and regularization coefficient $\beta$ from [0.005, 0.01, 0.05]. Dropout probabilities are fixed to 0.5. We also tune the layer number of noise filtering module from 1 to 3 and the layer number of causal self-attention module from 1 to 5.

We also implement all baselines expect PopRec, CLEA , GRec, FMLP-Rec and DualRec based on a comprehensive and efficient recommendation library RecBole~\cite{recbole}. 
For CLEA, GRec, FMLP-Rec and DualRec, we directly use their released code. For a fair comparison, we adopt the same data preprocessing, training and inference procedures as FMLP-Rec. We carefully tune the model hyper-parameters according to their original papers and report the best results.

\section{Additional Results of Performance Comparison in Full-Ranking Setting}\label{sec:frs_ana_app}

Following FMLP-Rec~\cite{fmlprec}, we select GRU4Rec, RepeatNet, HGN, SASRec, BERT4Rec, GCSAN, FMLP-Rec and DualRec as representative compared methods and HR@\{5, 10, 20\} and NDCG@\{5, 10, 20\} as metrics to further verify the effectiveness of \ours under full-ranking setting, which ranks the ground-truth item with all candidate items. Table~\ref{tab:per_com_full} shows the results, and we can find that \ours achieves better performance than other methods, which demonstrates the superiority of \ours.

 \begin{table*}[!t]
  \caption{Addition results of ablation study of \ours on ML100K, ML1M and Beauty datasets. Bold face indicates the highest performance.}
  \label{tab:abs_stu_supp}
  \resizebox{\linewidth}{!}{
  \small
  \begin{tabular}{l|ccc|ccc|ccc}
    \toprule
    &\multicolumn{3}{c|}{ML100K}&\multicolumn{3}{c|}{ML1M}&\multicolumn{3}{c}{Beauty}\\
    \midrule
    &HR@5&HR@10&NDCG@10&HR@5&HR@10&NDCG@10&HR@5&HR@10&NDCG@10\\
\midrule
(1) \ours w/o Noise Filtering Module&0.5073&0.6867&0.4136&0.7078&0.8083&0.5822&0.4042&0.5014&0.3418\\
(2) \ours w/ Learnable Filter&0.5143&0.6838&0.4183&0.6772&0.7803&0.5597&0.4052&0.5023&0.3435\\
(3) \ours w/o Future Information Encoder&0.4939&0.6768&0.3984&0.6839&0.7900&0.5621&0.4038&0.5054&0.3339\\
(4) \ours w/o Attenuation of Discrepancies&0.5107&0.6838&0.4164&0.6853&0.7939&0.5579&0.4269&0.5274&0.3582\\
(5) \ours w/ Traditional Training Strategy&0.5139&0.6880&0.4234&0.6877&0.7949&0.5651&0.4180&0.5191&0.3490\\
\midrule
(6) \ours w/ JS Diverge&0.5088&0.6789&0.4132&0.6999&0.8043&0.5739&0.4245&0.5224&0.3573\\
(7) \ours w/ Euclidean Distance&0.5128&0.6757&0.4151&0.7022&0.8077&0.5788&0.4204&0.5200&0.3516\\
(8) \ours w/ Cosine Distance&0.5224&0.6842&0.4234&0.7028&0.8060&0.5775&0.4224&0.5208&0.3528\\
\midrule
(9) \ours Future Information Encoder&0.4814&0.6513&0.3959&0.6629&0.7780&0.5438&0.3949&0.4894&0.3322\\
(10) \ours w/ R2L-style &0.5122&0.6800&0.4201&0.6689&0.7772&0.5510&0.4158&0.5161&0.3497\\
\midrule
(11) \ours (w/ KL Divergence \& L2R-style)&{\bf 0.5330}&{\bf 0.6908}&{\bf 0.4353}&{\bf 0.7106}&{\bf 0.8128}&{\bf 0.5863}&{\bf 0.4317}&{\bf 0.5295}&{\bf 0.3632}\\
    \bottomrule
  \end{tabular}
  }
\end{table*}

\section{Analysis of Loss in past information encoder}\label{sec:analysis_losses}
\begin{figure}[h!]
\centering
\includegraphics[width=1.0\linewidth]{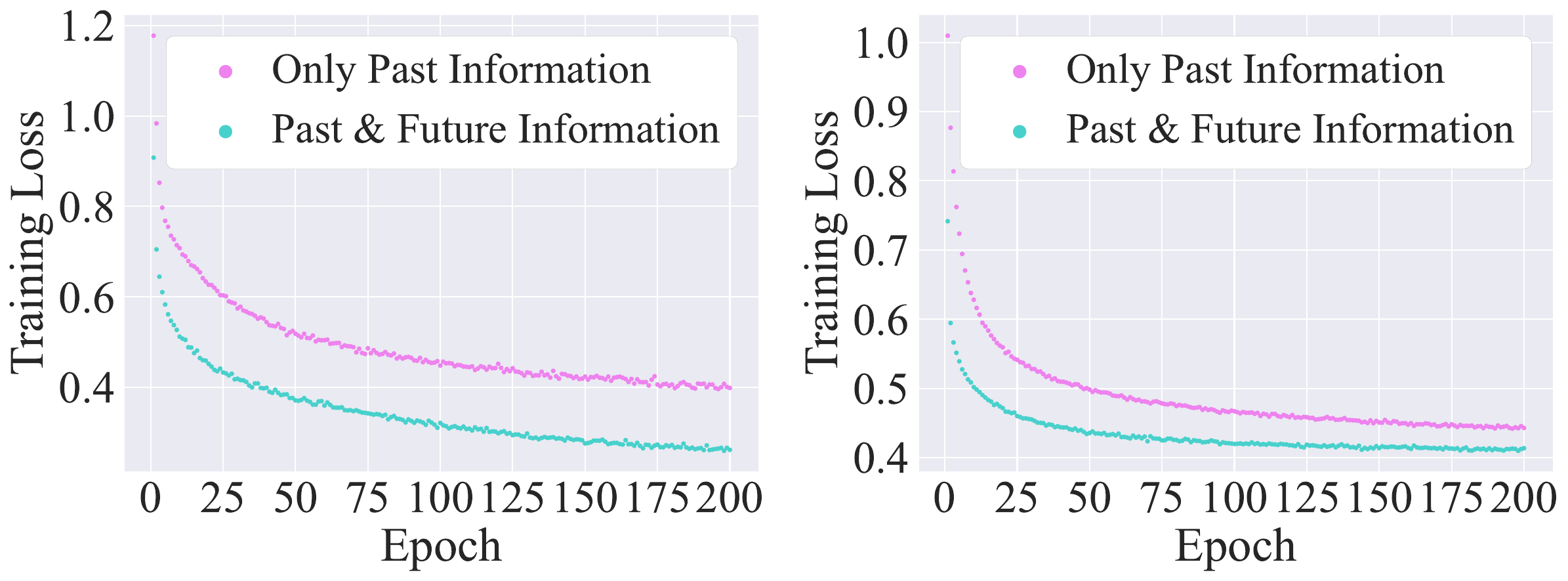}
\caption{The losses of past information encoder when using  past information only or past and future information together on ML100K (left) and ML1M (right) datasets.}
\label{fig:loss_ana}
\end{figure}

Figure~\ref{fig:loss_ana} shows the losses of past information encoder when using past information only or past and future information together to model user preferences on ML100K and ML1M datasets. We can find that using future information reduces the error of predicting user’s interested items during training, and thus will make more accurate recommendation in inference.

\section{Additional Results of Ablation Study}\label{sec:abs_stu_app}
Table~\ref{tab:abs_stu_supp} shows the additional results of ablation study of \ours on ML100K, ML1M and Beauty datasets. From the results, we can find that all components of \ours, i.e. noise filtering module, future information encoder, attenuation of discrepancies in oracle-guiding module, tailored 2PTraining and the left-to-right-style information extraction manner, have positive impact on the performance of \ours, and KL divergence is the most suitable discrepancy measurement method for \ours.

\section{Additional Results of Future Information Encoder Analysis}\label{sec:enc_ana_app}

\subsection{The Calculation of Preference Distribution}
We define user $u$'s preference distribution $p_u$ according to the categories (e.g., Comedy, Action) of items that are related to user $u$:
\begin{equation}
    p_{u}(\text{category}=c)=\frac{A_{uc}}{\sum_{c'=1}^CA_{c'k}},
\end{equation}
where $A_{uc}~(u=1,2,\cdots,M,~~~ c=1,2,\cdots, C)$ represents the number of appearances of the $c$-th category in the user $u$'s interacted items, $M$ is the number of users, and $C$ is the number of categories. There are three types of preference distributions used in the experiment:
\begin{enumerate}[1.]
    \item[$\bullet$] {\bf User historical preference distribution}, which is calculated from the categories of items that user has interacted with in the training phase.
    \item[$\bullet$] {\bf User real preference distribution}, which is calculated from the categories of items that user will interact with in the inference phase, which can reflect user real preference in the future.
    \item[$\bullet$] {\bf User predicted preference distribution}, which is calculated from the categories of top-10 items predicted by the model.
\end{enumerate}

\begin{figure*}[h!]
\centering
\resizebox{0.8\linewidth}{!}{
\includegraphics[width=1.0\linewidth]{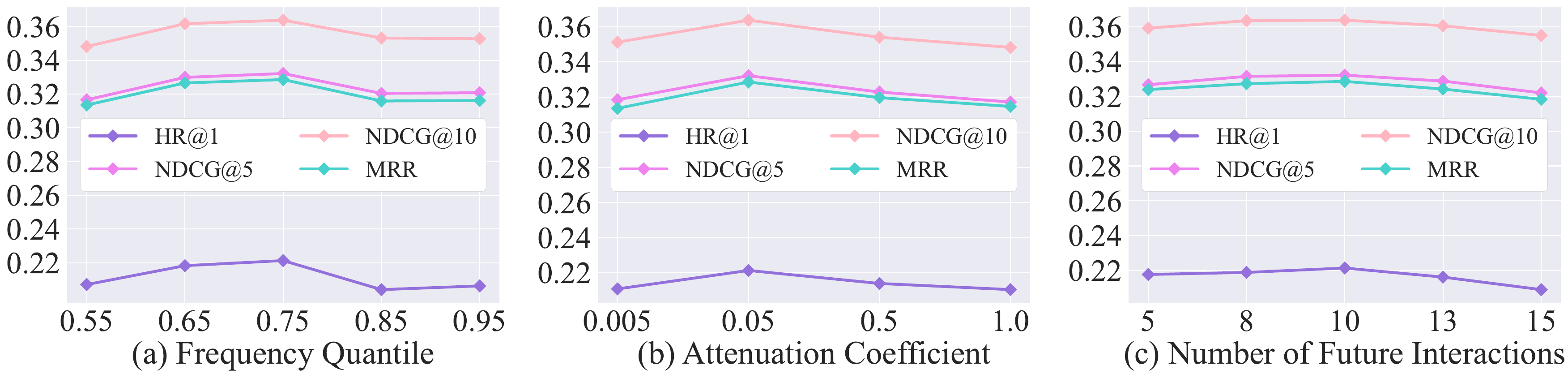}
}
\caption{Sensitivity analysis of three important hyper-parameters in \ours on Beauty.}
\label{fig:sen_ana_supp}
\end{figure*}

\subsection{The Calculation of KL Divergence}
KL divergence can be used to estimate the consistency of two distributions, thus we use KL divergence to evaluate the quality of the user predicted preference distribution with respect to user real preference distribution, where smaller KL divergence indicates better user predicted preference distribution. The KL divergence between user real preference distribution ($p$) and user predicted preference distribution ($q$) is:
\begin{equation}
    \mathbf{KL}(p,q) = \frac{1}{M}\sum_{u=1}^M\sum_{c=1}^Cp_u(c)\ln\frac{p_u(c)}{q_u(c)}.
\end{equation}

\subsection{Additional Cases of User Preference Distributions}

\begin{figure*}[t!]
\centering
\includegraphics[width=0.98\linewidth]{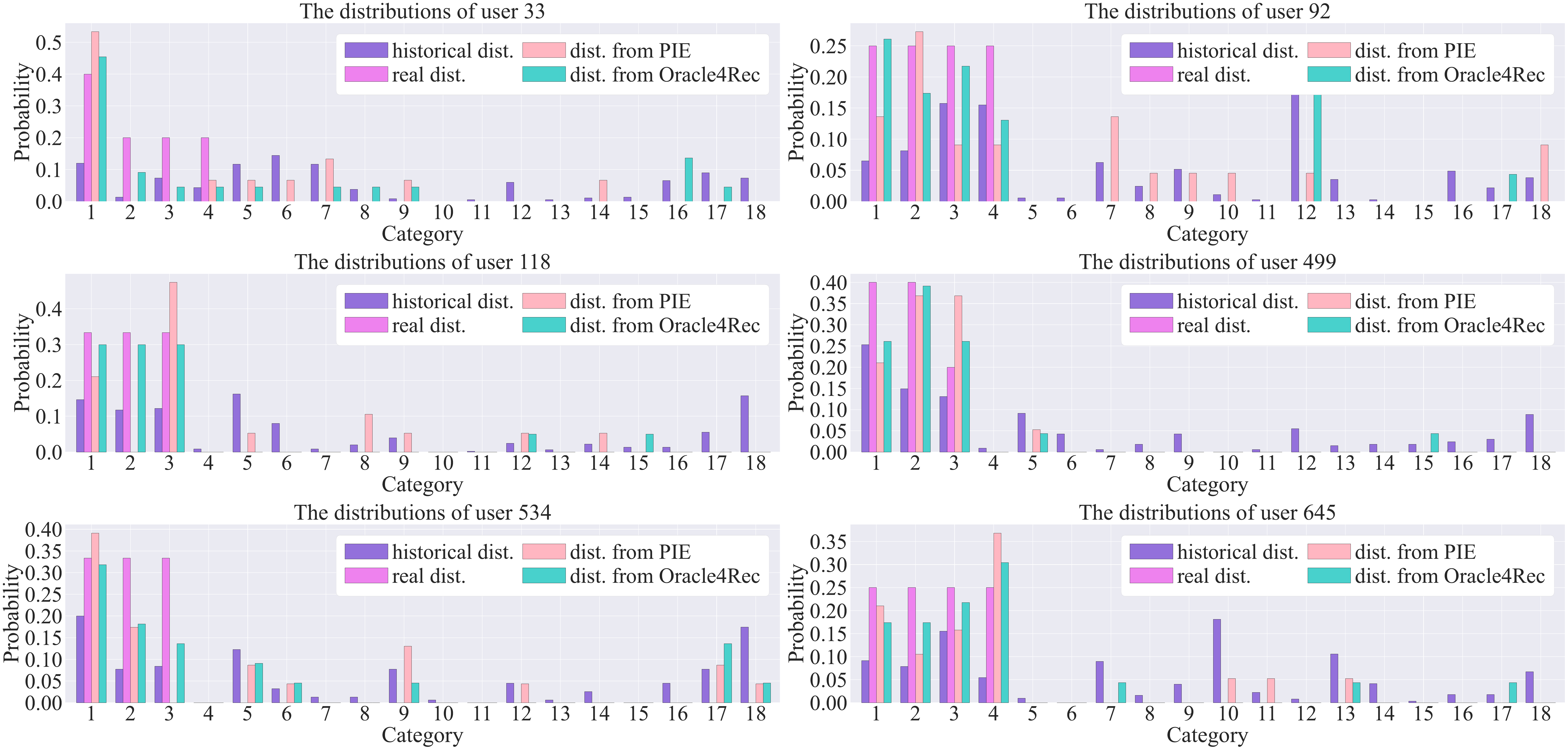}
\caption{Additional cases of four preference distributions of randomly selected 6 users on ML100K. ``dist.'' means distribution, and ``PIE'' is past information encoder, i.e., \ours w/o Future information Encoder.
}
\label{fig:case_supp1}
\end{figure*}

\begin{figure*}[t!]
\centering
\includegraphics[width=0.98\linewidth]{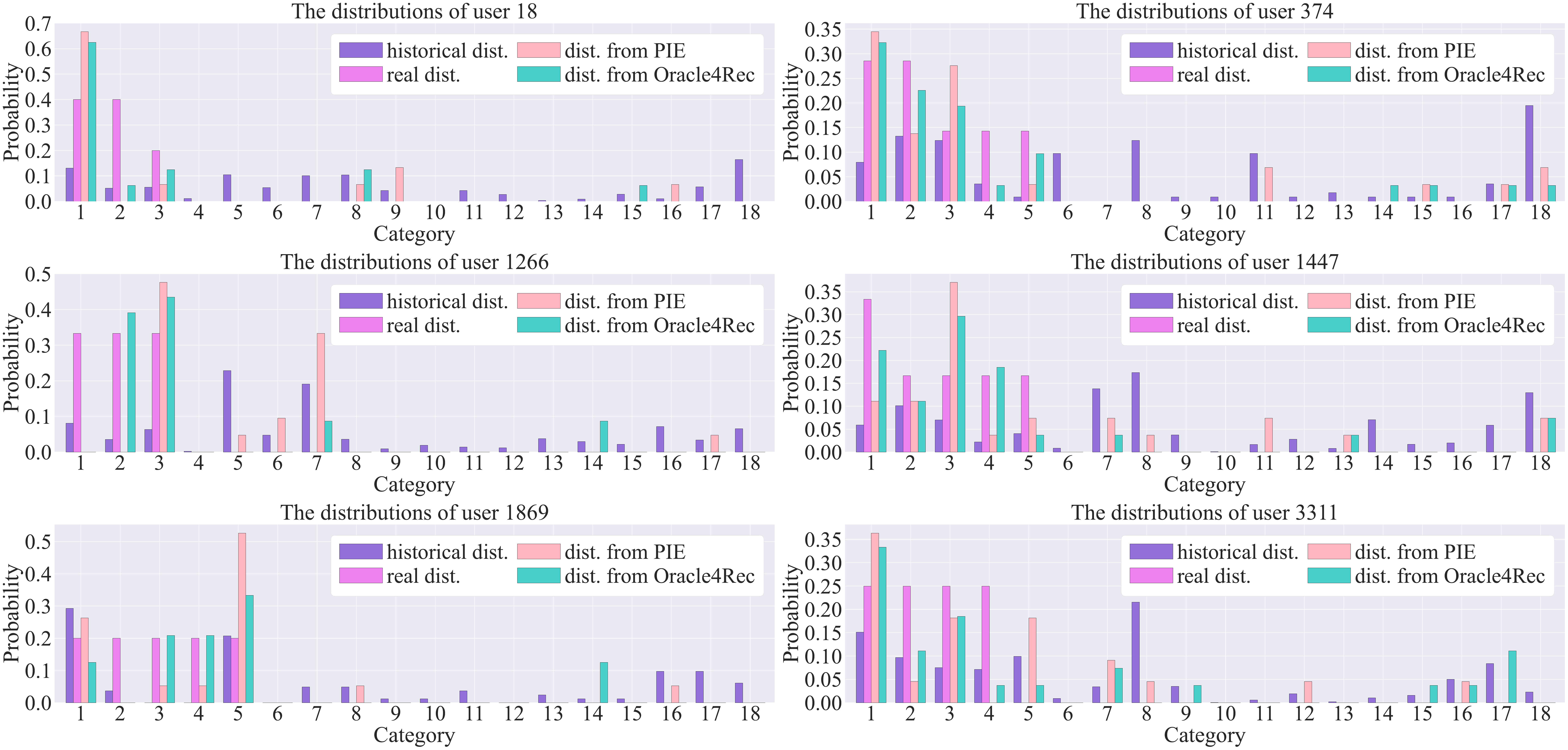}
\caption{Additional cases of four preference distributions of randomly selected 6 users on ML1M. ``dist.'' means distribution, and ``PIE'' is past information encoder, i.e., \ours w/o Future information Encoder.}
\label{fig:case_supp2}
\end{figure*}

Figure~\ref{fig:case_supp1} and Figure~\ref{fig:case_supp2} show the additional cases of four preference distributions of six randomly selected users on ML100K and ML1M respectively. From the results, we can find that merely using past information will make the model easily overfit historical preference. However, with the help of future information encoder, the probabilities of those categories are reduced by comparing distribution from PIE and distribution from \ours, alleviating the overfiting issue. Moreover, future information encoder makes the predicted preference distribution closer to the real distribution in inference phase by comparing real distribution, distribution from PIE and distribution from \ours, leading to more accurate user future interaction prediction.

\begin{table}[t!]
  \caption{Generality analysis of four sequential methods on ML100K and ML1M datasets. ``XYZ+'' means applying \ours into the corresponding method XYZ. Bold face indicates better performance in that group. The ``$\star$'' denotes the statistical significance ($p$ < 0.05) of the results of XYZ+ compared to XYZ.}
  \label{tab:gen_ana_supp}
  \resizebox{\linewidth}{!}{
  \begin{tabular}{l|ccc|ccc}
    \toprule
    &\multicolumn{3}{c|}{ML100K}&\multicolumn{3}{c}{ML1M}\\
    \midrule
    &HR@5&HR@10&NDCG@10&HR@5&HR@10&NDCG@10\\
\midrule
GRU4Rec&0.4795&0.6566&0.3872&0.6576&0.7754&0.5345\\
GRU4Rec+&{\bf 0.5116}$^\star$&{\bf 0.6793}$^\star$&{\bf 0.4181}$^\star$&{\bf 0.6800}$^\star$&{\bf 0.7901}$^\star$&{\bf 0.5602}$^\star$\\
\midrule
SRGNN&0.4806&0.6585&0.3946&0.6474&0.7567&0.5389\\
SRGNN+&{\bf 0.5060}$^\star$&{\bf 0.6744}$^\star$&{\bf 0.4119}$^\star$&{\bf 0.6708}$^\star$&{\bf 0.7826}$^\star$&{\bf 0.5606}$^\star$\\
\midrule
GCSAN&0.4993&0.6664&0.4121&0.6694&0.7773&0.5623\\
GCSAN+&{\bf 0.5228}$^\star$&{\bf 0.6859}$^\star$&{\bf 0.4267}$^\star$&{\bf 0.6864}$^\star$&{\bf 0.7913}$^\star$&{\bf 0.5803}$^\star$\\
\midrule
FMLP-Rec&0.5224&0.6783&0.4208&0.6830&0.7916&0.5648\\
FMLP-Rec+&{\bf 0.5275}&{\bf 0.6904}$^\star$&{\bf 0.4313}$^\star$&{\bf 0.6915}$^\star$&{\bf 0.7968}$^\star$&{\bf 0.5739}$^\star$\\
    \bottomrule
  \end{tabular}
 }
\end{table}

\section{Additional Results of Generality Analysis}\label{sec:gen_ana_app}

Table~\ref{tab:gen_ana_supp} shows the additional results of generality analysis on HR@5, HR@10 and NDCG@10. From the results, we can find that the performance of all methods has a significant improvement, which shows \ours has high generality.

\section{Results of Sensitivity Analysis}\label{sec:sen_ana_app}

We analyse how \ours performs with respect to three important hyper-parameters frequency quantile,  number of future interactions and attenuation coefficient on Beauty. Figure~\ref{fig:sen_ana_supp} shows the results. We neglect HR@5 and HR@10 since their range are different from others, but their trends are the same as others.

Figure~\ref{fig:sen_ana_supp}(a) shows the performance of \ours with frequency quantile varying in [0.55, 0.65, 0.75, 0.85, 0.95], and we can observe that when frequency quantile increases, the performance becomes better. This is because more frequency components are preserved, making the user preference modeling more precise. However, when frequency quantile is larger than 0.75, the performance is getting worse. This is because some high-frequency components, which correspond to the noise, are preserved, affecting the user preference modeling and model performance.

Figure~\ref{fig:sen_ana_supp} (b) shows the performance of \ours with attenuation coefficient varying in [0.005, 0.05, 0.5, 1.0]. We can find that with the increase of the attenuation coefficient, all metrics first increase and then decrease. When the attenuation coefficient is too small, \ours can not distinguish the importance of features in different periods, making it unable to learn the evolution of user preferences in the future. While when the attenuation coefficient is too large, \ours pays too much attention to the features in the near future, but ignores that of the longer term, making the user preferences modeling inadequate.

Figure~\ref{fig:sen_ana_supp} (c) shows the performance of \ours with the number of future interactions varying in [5, 8, 10, 13, 15]. From the results, we can observe that the best performance is achieved when it is 10. When it is too small, future information cannot be extracted from limited user future interactions, resulting in sub-optimal dynamic user preference modeling. When it is too large, excessive future interactions impact the model to extract effective and accurate information, and further impact the guiding of model training on past information, resulting in low performance.

\end{document}